\documentclass[twocolumn,iop,numberedappendix,twocolappendix]{aastex701}

\graphicspath{{./}{./figs/}}

\usepackage[T1]{fontenc}
\usepackage{newtxtext}
\usepackage[varg]{newtxmath} 

\usepackage{bm}
\usepackage{comment}
\usepackage{xspace}
\usepackage{amsmath}

\renewcommand{\eqref}[1]{Equation (\ref{#1})}
\renewcommand{\dfrac}[2]{ {\displaystyle\frac{#1}{#2}} }

\newcommand{\pr}[1]{\!\left(#1\right)}
\newcommand{\pd}[2]{\frac{\partial #1}{\partial #2}}

\newcommand{\avt}[1]{\ensuremath{\left\langle#1\right\rangle}\!}

\newcommand{\subtr}[1]{\Big[#1\Big]^{\rm bot}_{\rm top}}

\newcommand{\etaO}{\eta_{\rm O} }
\newcommand{\etaH}{\eta_{\rm H} }
\newcommand{\etaA}{\eta_{\rm A} }

\newcommand{\Am}{{\rm Am} }

\newcommand{\z}{$z$}
\renewcommand{\d}{{\rm d}}

\newcommand{\dz}{\,{\rm d}z}

\newcommand{\dr}{{\rm d}r}
\newcommand{\dth}{{\rm d}\theta}

\newcommand{\cs}{c_{\rm s}}

\newcommand{\rin}{r_\mathrm{in} }
\newcommand{\qrad}{q_{\rm rad} }

\newcommand{\qJ}{q_{\rm Joule} }
\newcommand{\qirr}{q_{\rm irr} }
\newcommand{\Tre}{T_{\rm re} }
\newcommand{\Tvis}{T_{\rm vis} }

\newcommand{\kapirr}{\kappa_{\rm irr} }
\newcommand{\kapd}{\kappa_{\rm disk} }
\newcommand{\Dth}{\delta }

\newcommand{\Trp}{T_{\rm r\phi}}
\newcommand{\Tpt}{T_{\rm \phi \theta}}

\newcommand{\tauxuv}{\tau_{r, \rm xuv}}
\newcommand{\tauirr}{\tau_{ r, \rm irr}}
\newcommand{\taudisk}{\tau_{\theta, \rm disk}}
\newcommand{\Lwind}{L_{\rm wind}}

\newcommand{\K}{\xspace\ensuremath{{\rm K}}\xspace}
\newcommand{\au}{\ensuremath{{\rm au}}\xspace}
\newcommand{\yr}{\ensuremath{\rm \,yr}\xspace}
\newcommand{\Msun}{\ensuremath{\,M_{\sun}}\xspace}
\newcommand{\Lsun}{\ensuremath{\,L_{\sun}}\xspace}
\newcommand{\Mpyr}{\xspace\ensuremath{\Msun \,\rm \yr^{-1}}\xspace}
\newcommand{\cmcmg}{\ensuremath{\rm cm^{2} \, g^{-1}}\xspace}
\newcommand{\gcmcm}{\xspace{\ensuremath{\rm g \, cm^{-2}}}\xspace}
\newcommand{\ergs}{\xspace{\ensuremath{\rm erg \, s^{-1}}}\xspace}

\newcommand{\rFHp}{\texttt{FidH+}\xspace}
\newcommand{\rFHn}{\texttt{FidH-}\xspace}%
\newcommand{\rFNH}{\texttt{FidNoH}\xspace}
\newcommand{\rOHp}{\texttt{OpH+}\xspace}
\newcommand{\rOHn}{\texttt{OpH-}\xspace}
\newcommand{\rONH}{\texttt{OpNoH}\xspace}

\defcitealias{Chiang1997Spectral-Energy}{CG97}
\defcitealias{Hubeny1990Vertical-struct}{H90}
\defcitealias{Bai2017Global-Simulati}{B17}

\newcommand{\Hini}{H_{\rm mid, 0} }
\newcommand{\Hmid}{H_{\rm mid} }
\newcommand{\rhomini}{\rho_{\rm mid, 0} }

\newcommand{\Tmid}{T_{\rm mid} }
\newcommand{\djdr}{j_0' }
\newcommand{\sint}{\sin\mspace{-2mu}\theta\mspace{2mu}}
\newcommand{\nablap}{\nabla_{\rm \!p}\!}

\begin{document}

\title{Radiative Nonideal MHD Simulations of Inner Protoplanetary Disks: Temperature Structures, Asymmetric Winds, and Episodic Surface Accretion}

\correspondingauthor{Shoji Mori}

\def\tsinghua{Institute for Advanced Study and Department of Astronomy, Tsinghua University, Beijing 100084, China}

\author[0000-0002-7002-939X]{Shoji Mori}
\affiliation{Astronomical Institute, Tohoku University, 6-3 Aramaki, Aoba-ku, Sendai 980-8578, Japan}
\affiliation{\tsinghua}
\email{mori.s@astr.tohoku.ac.jp}

\author[0000-0001-6906-9549]{Xue-Ning Bai}
\affiliation{\tsinghua}
\email{}

\author[0000-0001-8105-8113]{Kengo Tomida}
\affiliation{Astronomical Institute, Tohoku University, 6-3 Aramaki, Aoba-ku, Sendai 980-8578, Japan}
\email{}

\begin{abstract}
We perform two-dimensional global magnetohydrodynamic (MHD) simulations including the full nonideal MHD effects (Ohmic diffusion, Hall effect, and ambipolar diffusion) and approximate radiation transport to understand the dynamics and thermal structure of the inner protoplanetary disks (PPDs). 
We have developed a simple radiative transfer model for PPDs that reasonably treats stellar non-thermal (XUV), stellar thermal (optical/infrared), and re-emitted radiations, reproducing the temperature structures from Monte Carlo radiative transfer. 
Our simulations show fast one-sided surface accretion ($\sim 10\%$ of Keplerian velocity) and asymmetric disk winds when the vertical magnetic field is aligned with the disk angular momentum. 
The asymmetry is due to the failure of the wind on the side with the accretion layer. 
On the accreting surface, clumps are repeatedly generated and accrete, driven by radiative feedback. 
For the anti-aligned fields, surface accretion becomes more moderate and time-variable, while the winds remain largely symmetric. 
For the thermal structure, accretion heating does not affect the disk temperature in any of our runs. 
This is because (1) the accretion energy dissipates via Joule heating at 2--3 gas scale heights, where low optical depth enables efficient radiative cooling, and (2) the winds remove $\gtrsim 10\%$ of the accretion energy. 
In contrast, the winds enhance radiative heating by elevating the irradiation front. 
These results highlight the importance of coupling between gas dynamics and radiation transport in PPDs,
and provide observable magnetic activities such as fast episodic accretion, wind asymmetry, and molecular survival in XUV-irradiated winds.

\end{abstract}



\section{Introduction}\label{sec:intro}

Understanding the physical structure of the inner 10 au of protoplanetary disks (PPDs) is the key to understanding planet formation and migration, especially with respect to the exoplanet population.
In particular, the temperature structure is fundamental to many of the planet formation processes.
For instance, the disk temperature determines the location of the snowlines of water and other volatiles, thus controlling the main composition of dust.
The snowline is also expected to be the preferred location for the formation of planetesimals and gas planets 
\citep{Stevenson1988aRapid-formation,Ros2013Ice-condensatio,Drazkowska2017aPlanetesimal-fo,Schoonenberg2019Pebble-driven-p}.
The disk temperature also controls the chemistry, thereby influencing the atmospheric composition of planets \citep{Dawson2018Origins-of-Hot-,Oberg2021Astrochemistry-}.
Furthermore, various planet formation processes, such as pebble isolation and planetary migration, depend on the disk temperature \citep[e.g.,][]{Paardekooper2009On-corotation-t,Lambrechts2014Separating-gas-,Bitsch2018Pebble-isolatio,Johansen2019How-planetary-g,Liu2020A-tale-of-plane}.
Therefore, understanding the disk temperature is essential to accurately model the planet formation processes.
Recently, JWST mid-infrared observations are revealing the physical and chemical structures of inner PPDs
\citep[e.g.,][]{Xie2023Water-rich-Disk,Gasman2023MINDS.-Abundant,Schwarz2024MINDS.-JWST/MIR}.
A better understanding of the physical structure of the inner disk region is essential to properly interpret these observations.








 
The temperature structure of the inner PPDs is mainly determined by irradiation heating and accretion heating. 
Irradiation heating is the mechanism by which stellar radiation illuminates the disk surface, and the re-emitted radiation from the heated surface warms the disk interior (\citealt[][hereafter \citetalias{Chiang1997Spectral-Energy}]{Kenyon1987Spectral-energy,Calvet1991Irradiation-of-,Chiang1997Spectral-Energy}).
Accretion heating is powered by the release of gravitational potential energy,
mediated by internal dissipation \citep{Lynden-Bell1974The-evolution-o,Balbus1999On-the-Dynamica}.

Classical models of viscous accretion disks generally rely on the presence of turbulence especially from the magnetorotational instability (MRI) 
to drive outward angular momentum transport \citep{Balbus1991A-powerful-loca,Hawley1995Local-Three-dim}.
Such turbulence also leads to the dissipation of accretion energy, known as viscous heating.
Viscous heating primarily releases the energy around the (generally) optically thick midplane, where the density is higher, and
can thus efficiently warm the disk.

However,
PPDs are very weakly ionized, leading to three non-ideal MHD effects including
Ohmic diffusion, ambipolar diffusion, and the Hall effect \citep[e.g.,][]{Sano2000Magnetorotation,Wardle2007Magnetic-fields,Turner2014Transport-and-A}.
The Ohmic and ambipolar diffusion cause dissipation of magnetic fields and dominate in high-density and low-density regions, respectively. These effects suppress or damp the MRI turbulence \citep{Bai2013aWind-driven-Acc,Bai2013bWind-driven-Acc,Gressel2015Global-Simulati}.
In the inner $\sim10$ au region of PPDs in particular, the disk is expected to be largely laminar. Driving disk evolution that matches the observed accretion rates requires the presence of net poloidal magnetic fields threading the disk, launching magnetized disk winds
\citep{Bai2013aWind-driven-Acc,Bai2017Hall-Effect-Med,Gressel2015Global-Simulati}.

In addition, the magnetic field structure is strongly affected by the Hall effect in PPDs, depending on the polarity of the magnetic field threading the disk. 
The Hall effect dominates at intermediate density and magnetic field strengths (in between Ohmic and ambipolar diffusion), and leads to drifts of magnetic flux in the direction of the electric current.
When the magnetic field is aligned with the disk's angular momentum, the Hall effect amplifies the radial and toroidal fields \citep[Hall-shear instability, HSI;][]{Kunz2008On-the-linear-s,Bai2014Hall-effect-Con,Lesur2014Thanatology-in-}.
On the other hand, for the anti-aligned field polarity, the Hall effect weakens the radial and toroidal fields.
It is important to consider these nonideal MHD effects for understanding magnetically controlled PPDs.

Our research aims to understand the heating processes in the largely laminar magnetized PPDs.
In our previous study \citep{Mori2019Temperature-Str}, we performed MHD simulations incorporating the full nonideal MHD effects for the inner region of PPDs, using a local shearing box. 
These simulations revealed that heating due to magnetic diffusion (a.k.a. Joule heating) occurs predominantly at high altitudes rather than at the midplane, making it relatively inefficient \citep[see also][]{Hirose2011Heating-and-Coo,Bethune2020Electric-heatin}. 
\citet{Mori2021Evolution-of-th} modeled the heating altitude from \citet{Mori2019Temperature-Str} and demonstrated its impact on disk temperature and the location of the snowline.
\citet{Kondo2023The-Roles-of-Du} further demonstrated how Joule heating is affected by dust growth, which influences snowline evolution.
\citet{Mori2025Long-term-evolu} showed that mass loss due to disk winds decreases the accretion rate toward the inner disk, which in turn reduces the accretion heating.

However, several questions remain as follows.
\begin{itemize}
\item
{\it Heating due to the global-scale magnetic field.}
Local MHD simulations of inner regions of PPDs 
likely underestimate Joule heating. A physical magnetic configuration demands horizontal field lines to flip between the upper/lower surface, which is usually not present in shearing box simulations due to its intrinsic symmetry \citep{Bai2013aWind-driven-Acc}.
Moreover, the radial gradient in toroidal magnetic fields is missing in the shearing box, which may lead to additional Joule heating in the disk interior.   


\item
{\it Wind properties.}
The wind properties should be determined by the global-scale magnetic field configuration that is not captured in local models. 
In particular, shearing box simulations tend to overestimate the wind mass loss rate due to artificial truncation at the vertical boundary \citep{Fromang2013Local-outflows-,Bai2013aWind-driven-Acc}. This would also affect the energy budget in the wind.

%
%
%
%
%

\item
{\it Gas dynamics consistent with the thermal structure.}
Our previous studies relied on prescribed thermodynamics
(\citealt[][hereafter \citetalias{Bai2017Global-Simulati};]{Bai2017Global-Simulati} \citealt{Mori2019Temperature-Str}). 
However, the thermal structure influences gas dynamics and wind properties \citep{Bai2016Magneto-thermal,Gressel2020Global-Hydromag},
calling for consistent treatment of both MHD and thermodynamics for a proper understanding of PPD gas dynamics.
\end{itemize}

In this study, we perform non-ideal MHD simulation (including the Hall effect) with approximate radiation transport in a two-dimensional global computational domain.
The global domain allows us to eliminate the concerns left over from the local simulation \citep{Mori2019Temperature-Str}.
In addition, by taking into account the radiation transport, 
we can obtain the temperature structure and evolution consistent with the disk dynamics.
Furthermore, we highlight differences in the disk structures depending on the polarity of the poloidal magnetic field, which is caused by the Hall effect.



This paper is organized as follows. In Section~\ref{sec:method}, we outline our simulation methodology, detailing the radiation transport scheme and simulation setup. 
Section~\ref{sec:results} presents the disk dynamics and temperature structures obtained in the simulations, varying the magnetic field polarity and optical opacity. 
Section~\ref{sec:discussion} compares these results with previous temperature models and observations.
Finally, we summarize our findings in Section~\ref{sec:sum}.

\section{Method} \label{sec:method}
The basic equations for the MHD simulations are described in Section~\ref{ssec:method-basic},
while the method to solve the radiative transfer is described in Section~\ref{ssec:method-rt}.
Section~\ref{ssec:simulation-setup} presents the simulation setup.

\subsection{Basic Equations} \label{ssec:method-basic}

For the calculation of MHD, we follow the method of \citetalias{Bai2017Global-Simulati}. 
We solve MHD equations with Athena++ \citep{Bai2017Hall-Effect-Med,Stone2020The-Athena-Adap}:
\begin{eqnarray}
    &&\pd{\rho}{t} + \nabla \cdot \pr{ \rho \bm{v} } = 0  ,\\
    &&\pd{ \pr{ \rho \bm{v} } }{t} + \nabla \cdot \pr{\rho \bm{v} \bm{v}  - \frac{\bm{B} \bm{B}}{4 \pi} +  P^* \mathsf{I}} = - \rho \nabla \Phi ,\\
    &&\pd{\bm{B}}{t} = \nabla \times \pr{ \bm{v} \times \bm{B}} -c \nabla \times \bm{\mathcal{E}}' ,  \\
    &&\pd{(E + \rho \Phi)}{t} + \nabla \cdot \left [ \pr{E + \rho \Phi + P^*}\bm{v} \right . \nonumber   \\
    && \left. ~~~~ - \bm{B} \pr{\bm{B} \cdot \bm{v} } / (4 \pi) + \bm{S} \right] =   \qrad  \label{eq:ene-eq},
\end{eqnarray}
where
$\rho$ is the density, 
$\bm{v}$ is the gas velocity, 
$\bm{B}$ is the magnetic field, 
$P^*$ is the sum of gas pressure $P$ and magnetic pressure $B^2/8\pi$, 
$\mathsf{I}$ is the identity tensor,
$\Phi = - G M_\odot/r$ is the gravitational potential of a point source with 1 $M_\odot$,
$\bm{\mathcal{E}}'$ is the electric field of the non-ideal MHD effects in the frame co-moving with the gas, 
$E$ is the total energy density, 
$\bm{S}$ is the Poynting vector $c \bm{\mathcal{E}}' \times \bm{B}$,
and $\qrad$ is the source term from radiative transfer (see Section~\ref{ssec:method-rt}).
$\bm{\mathcal{E}}'$ is given by 
\begin{equation}
	\bm{\mathcal{E}}' = \frac{4 \pi}{c^2} \pr{ \etaO \bm{J} + \etaH \bm{J} \times \bm{b} + \etaA \bm{J_\perp} } ,
\end{equation}
where $\bm{J}$ is the electric current density $\bm{J} = (c/4\pi) \nabla \times \bm{B}$ with
$\bm{J}_\perp=- (\bm{J} \times \bm{b} ) \times \bm{b} $ being its component perpendicular to $\bm{B}$, 
and $\bm{b}$ is the unit vector along $\bm{B}$.
The diffusivities for Ohmic, Hall, and ambipolar diffusion are denoted by
$\etaO$, $\etaH$, and $\etaA$, respectively.
The total energy density is $E =  \rho \bm{v}^2 /2 + \bm{B}^2/8\pi +  \rho e $, where $\rho e$ is the internal energy density with the adiabatic index of 7/5. 
The MHD equations are solved by the HLLE Riemann solver \citep{Einfeldt1991On-Godunov-Type}, which is necessary for implementing the Hall effect \citep{Lesur2014Thanatology-in-}, with the constrained transport method \citep{Evans1988Simulation-of-M}.
We adopt the second-order piece-wise linear spatial reconstruction and van-Leer time integrator, 
and the Ohmic and ambipolar diffusion terms are solved by operator splitting with super timestepping \citep{Meyer2014A-stabilized-Ru}. 
We also set a diffusivity cap of $10 \Hini^2 \Omega$ to ensure that the simulation timestep does not become prohibitively small, where $\Hini$ is the initial midplane gas scale height (see Section~\ref{ssec:simulation-setup}) and $\Omega$ is the Keplerian angular velocity.


We carry out two-dimensional (axis-symmetric) global simulations in spherical-polar coordinates ($r$, $\theta$). For convenience, cylindrical coordinates ($R$, $z$) 
and the latitude $\Dth = \pi/2 - \theta$ 
are also used in setting the initial conditions and in the analysis.

%

The non-ideal MHD diffusivities are obtained by referencing a diffusivity table given as a function of local gas density $\rho$, $T$, and the ionization rate. We adopt the table used in \citetalias{Bai2017Global-Simulati}, following the evolution of a complex chemical network \citep{Bai2009Heat-and-Dust-i,Bai2011The-Role-of-Tin} for $10^6$ years to equilibrium, assuming a grain size of $0.1 {\rm \mu m}$ and a dust mass fraction of $10^{-4}$. 
The adopted dust mass fraction reflects reduced total surface area compared to interstellar medium dust, e.g., due to grain growth (up to $\sim$1 mm) and settling \citep[e.g.,][]{Birnstiel2011Dust-size-distr}.
The scaling of the magnetic field strength for the diffusivities is assumed to be $\etaO \propto B^0 $, $\etaH \propto B^1 $, and $\etaA \propto B^2$, as justified in \citet{Xu2016On-the-Grain-mo}.


To calculate the ionization rate, we follow \citetalias{Bai2017Global-Simulati} and consider ionization due to cosmic rays, stellar X-rays, stellar far UV (FUV), and short-lived radionuclides. 
These ionization rates are based on the column densities along the $r$-direction ($\Sigma_r$) and $\theta$-direction ($\Sigma_\theta$),
\begin{eqnarray}
	\Sigma_r (r, \theta) &=& \int^r_{\rin} \rho(r, \theta) \dr + \rho(\rin, \theta) \rin,  \label{eq:Sigma-r} \\
	\Sigma_\theta (r, \theta) &=& \min\left[  \int_{\theta_{\rm s}}^\theta \rho(r, \theta) r \dth ,  \int_{\theta}^{\theta_{\rm e}} \rho(r, \theta) r \dth  \right] ,
\end{eqnarray}
where $\rin$ is the inner edge radius of the simulation domain, and we set $\theta_{\rm s} = \pi/6$ and $\theta_{\rm e} = 5\pi/6$ to exclude the polar region.
The ionization rate for cosmic rays is given by $10^{-17} {\rm s}^{-1} \exp\pr{-\Sigma_\theta / 96\, \gcmcm} $ \citep{Umebayashi1981Fluxes-of-Energ}.
For X-rays, we give the ionization rate by the sum of the direct component (dependent on $\Sigma_r$) and scattered component (dependent on $\Sigma_\theta$), with an X-ray luminosity of $10^{30} \ergs$ and an X-ray temperature of 3 keV using the fitting formula of \citet{Bai2009Heat-and-Dust-i} based on \citet{Igea1999X-Ray-Ionizatio}.
Ionization from short-lived radionuclides is given by a constant rate of 6.0 $\times$ 10$^{-19}$ s$^{-1}$ \citep{Umebayashi2009Effects-of-Radi}.
FUV ionization is treated in a special manner motivated by \citet{Perez-Becker2011Surface-Layer-A}, which penetrates the disk up to a surface density of $\Sigma_{\rm FUV}\sim$ 0.01--0.1\,\cmcmg that nearly fully ionizes C and S followed by a sharp cutoff. Following \citetalias{Bai2017Global-Simulati}, we overwrite the diffusivities when they are smaller than those due to the FUV ionization fraction,
\begin{eqnarray}\label{eq:fuv}
	x_{\rm FUV} = 2 \times 10^{-5} \max\pr{g, 1} \exp  \pr{  - \tauxuv^4  } ,
\end{eqnarray}
where 
$g = \exp[0.3/(\tauxuv + 0.05)]$ is an artificial factor boosting the ionization fraction in lower-density regions to ensure the wind gas behaves as ideal MHD (thanks to additional ionization processes by extreme UV and/or soft X-rays), and 
$\tauxuv = \kappa_{\rm xuv} \Sigma_r$
is the effective optical depths of such ionizing radiation from the star, with $\kappa_{\rm xuv}$ taken to be $\Sigma_{\rm FUV}^{-1}$. 
%

\subsection{Simplified Calculation of Radiation Transport}
\label{ssec:method-rt}


We calculate the radiative heating source term $\qrad$ using an approximate method inspired by \citet{Xu2021Formation-and-e},
which captures optically-thin cooling and optically-thick diffusion with a smooth transition in between.
The radiative heating source term is given by
\begin{equation}
\begin{split}\label{eq:Qrad}
    \qrad = &\exp\pr{-\taudisk} \pr{q_{\rm h,thin} - q_{\rm c,thin}} \\
    &- \nabla \cdot \left[ \exp\pr{-1/\taudisk} \bm{F}_{\rm thick} \right] ,
\end{split}
\end{equation}
where 
\begin{equation}
	\taudisk =\kappa_{\rm disk} \Sigma_\theta
\end{equation}
represents the optical depth for disk radiation in the $\theta$ direction,
$q_{\rm h,thin}$ and $q_{\rm c,thin}$ are the heating rate and cooling rate per volume (applicable in the optically thin regime), and $\bm{F}_{\rm thick}$ is the energy flux from radiative diffusion (applicable in the optically thick regime).

The heating rate $q_{\rm h,thin}$ considers the X-ray and FUV heating (XUV), stellar irradiation (optical/infrared), and the re-emitted radiation from the disk,
\begin{equation}
\begin{split}\label{eq:thin-heat}
       q_{\rm h,thin} = 4 \rho \sigma \Bigl[
			 &\kappa_{\rm xuv} T_{\rm xuv}^4 \exp\pr{- \tauxuv^4 } \\
    			 + &\kapirr T_{\rm irr}^4 \exp\pr{- \tauirr} 
			+ \kappa_{\rm disk} T_{\rm re}^4 \Bigr],
\end{split}
\end{equation}
where
$\sigma$ is the Stefan-Boltzmann constant,
$\kappa_{\rm xuv}$, $\kapirr$, and $\kappa_{\rm disk}$ are the effective opacity for the XUV heating, stellar radiation, and disk emission, respectively,
and 
$\tauirr = \kapirr \Sigma_r$ is the irradiation optical depth from the star.
We note that our prescription for XUV heating is artificial with a functional form set in accordance with that for FUV ionization. More realistic calculations would involve many photo-chemical processes and the associated heating and cooling, which is beyond the focus of this work, and the readers may refer to \citet{Wang2019Global-Simulati}, and \citet{Gressel2020Global-Hydromag}.
We give the radiation temperature for the thermal stellar radiation and XUV radiation as,
\begin{eqnarray}
	T_{\rm irr, eq}(r) &=& 360 \,\K ~(r/\au)^{-0.5} , \label{eq:Tirr} \\
	T_{\rm xuv, eq}(r) &=& 3600 \,\K ~(r/\au)^{-0.5} , \label{eq:Txuv}
\end{eqnarray}
where $T_{\rm irr, eq}$ is related to the stellar luminosity $L$ (taken to be $L \approx 2.7 L_\odot$) as $T_{\rm irr, eq}^4 = L/(16 \pi r^2 \sigma)$ \footnote{This temperature is given from the thermal balance far from the star, $ \rho \kappa (u - 4  \sigma T^4 /c ) = 0 $, with the radiation energy density $u$ being $u \approx L / (4 \pi r^2 c)$ and $c$ being the light speed.},
and we mimic the XUV heating by setting $T_{\rm xuv, eq}$ to be 10 times higher than $T_{\rm irr, eq}$ \citep[e.g.,][]{Nakatani2018aRadiation-Hydro}.
The radiation temperature $T_{\rm re}$ for the re-emitted radiation from dust is calculated along the $\theta$-direction by solving the one-dimensional radiative transfer equation under the two-stream approximation (see Appendix~\ref{app:Tre}).
The cooling rate is given by 
\begin{eqnarray}
       q_{\rm c, thin} &=& 4 \rho \kappa_{\rm disk} \sigma T^4 . 
\end{eqnarray}
Our fiducial choice of $\kappa_{\rm disk}$, $\kapirr$ and $\kappa_{\rm xuv}$ are listed in Table \ref{tab:parameters}. As an initial effort, we adopt constant opacities for simplicity.
Also, we assume the gas is cooled by its own temperature, though the dust-gas mixture should be cooled by the dust emission.
This approach is, however, a reasonable approximation and it ensures that the temperature converges well to the target temperature.

In the optically thick region, the radiative energy flux $ \bm{F}_{\rm thick} $ follows 
the diffusion approximation: $ \bm{F}_{\rm thick} = - (4 \sigma/3\rho \kappa_{\rm disk} )  \nabla T^4 $.
The optically-thick term in Equation (\ref{eq:Qrad}) is solved as thermal conduction,
\begin{eqnarray}
&&  - \nabla \cdot \left[ \exp\pr{-1/\taudisk} \bm{F}_{\rm thick} \right] 
    = \nabla \cdot \left[ k_{\rm diff} \rho \nabla \cs^2 \right ],\\
    &&k_{\rm diff} = \exp\pr{-1/\taudisk} \frac{16\sigma T^4 }{3 \rho^2 \kappa_{\rm disk} \cs^2}  ,
\end{eqnarray}
where $k_{\rm diff}$ is the diffusivity for the radiative diffusion and includes the cutoff function, 
$\cs(T) = \sqrt{k_{\rm B} T/ m_{\rm g} }$ is the sound speed, 
$k_{\rm B}$ is the Boltzmann constant,
and $m_{\rm g}$ is the mean gas particle mass (2.3 amu).
Because of the strong temperature dependence in $k_{\rm diff}$, we find that using operator splitting with subcycling or super timestepping will render the scheme unstable. To mitigate the numerical timestep,
we employ a cap to $ k_{\rm diff} $ with $100 \Hini^2 \Omega$.
We have confirmed that thanks to our approach to smoothly transition from optically thick to optically thin regimes, the simulation results remain unchanged as long as the cap is above $ 10 \Hini^2 \Omega$.

We have validated this approach of approximate radiation transport through comparison with a Monte Carlo simulation and confirmed it has achieved the desired accuracy (see Appendix~\ref{app:test}). 


\subsection{Simulation Setup} 
\label{ssec:simulation-setup}

\begin{deluxetable}{ll}
\tablecaption{Fiducial Simulation Parameter Set \label{tab:parameters}}
\tablehead{
\colhead{Parameter} & \colhead{Value} 
}
\startdata
Radial computational domain & [0.4, 40] au \\
Stellar mass, $M_*$                          & 1  $M_\odot$  \\
Stellar luminosity 		   & 2.7  $L_\odot$  \\
Disk surface density at 1 au & 320 $\gcmcm$ \\
Slope of initial surface density & $-1$   \\
Initial midplane temperature at 1 au & 150 $\K$ \\
Slope of initial temperature & $-0.5$   \\
Initial midplane gas-to-magnetic pressure, $\beta_{\rm mid, 0}$ & $10^4$   \\
Effective opacity for X-rays and FUV, $\kappa_{\rm xuv}$ & 33  \cmcmg \\
Opacity for stellar irradiation, $\kapirr$ & 1 \cmcmg  \\
Opacity for disk radiation, $\kappa_{\rm disk}$ & 1 \cmcmg \\
Radiation temperature of XUV irradiation at 1 au  & $ 3600 $ K  \\
Radiation temperature of optical irradiation at 1 au  & $ 360 $  K  \\
Ionization rate of cosmic rays & 10$^{-17}$ ${\rm s}^{-1}$ \\
Luminosity of stellar X-ray & $10^{30}$ \ergs \\ 
\enddata

\end{deluxetable}

Table \ref{tab:parameters} summarizes the fiducial parameter values in our simulations.
The simulation domain is set with $r = [0.4, 40] $ au and $\theta = [0, \pi]$, with a grid resolution of (512, 320). We use logarithmic spacing in $r$, while grid spacing
along the $\theta$-direction is non-uniform, as in \citetalias{Bai2017Global-Simulati}, with $\Delta\theta$
$\approx 0.0015$ around the midplane, and $\Delta\theta \approx 0.005$ around the disk surface ($\delta \sim 0.2$).

In our simulations, we add physical effects in steps to allow smooth evolution toward quasi-steady states.
At the start of the calculation ($t = 0$), neither magnetic field nor radiative transfer is employed.
At $t = 0.1$ yr, we apply a poloidal magnetic field  while also turning on Ohmic and ambipolar diffusion.
At $t = 150$ yr, we enable the Hall effect (unless in runs where we exclude the Hall effect) by increasing $\etaH$ linearly over 50 yrs so that the HSI can develop gradually (see Appendix A of \citetalias{Bai2017Global-Simulati} for further discussion).
When the Hall effect is fully activated ($t = 200$ yr), we switch on the radiation transport.

At the beginning of the simulations, the initial temperature distribution is given by:
\begin{equation}
\begin{split}
\label{eq:Tini}
	T =&  \max(T_{\rm disk,0},  T_{\rm irr, 0}, T_{\rm xuv, 0})
\end{split} 
\end{equation}
where
\begin{eqnarray}
    T_{\rm disk, 0} (R) &=& 150 \,\K ~(R/\au)^{-0.5} , \\
    T_{\rm irr, 0}(r, \delta) &=& T_{\rm irr,eq}(r) f_{\rm cut}(\Dth/3h_0) , \\
    T_{\rm xuv, 0}(r, \delta) &=& T_{\rm xuv,eq}(r) f_{\rm cut}(\Dth/9h_0) , \\
    f_{\rm cut}(x) &=& \max\left[1 - 1/x^8 , 0 \right],
\end{eqnarray}
and 
the initial gas scale height is $\Hini = \cs(T_{\rm disk, 0})/\Omega \propto R^{1.25}$,
and $h_0 = \Hini/R$ is its aspect ratio.
The subscript 0 represents the value at the beginning of the simulation.
The second and third terms of Equation (\ref{eq:Tini}) mimic the temperature structure due to stellar irradiation and XUV-heating, as in Equations (\ref{eq:Tirr})--(\ref{eq:Txuv}).
The obtained quasi-steady states are not sensitive to the initial temperature structure.
We enforce the initial temperature structure until switching on the radiation transport at $t = 200$ yr.

The initial surface density profile is given by 
\begin{equation}\label{eq:Sigma}
	\Sigma_0 = 320 \gcmcm \times \pr{R/\au}^{-1} .
\end{equation}
The density distribution is given as in the vertically-isothermal hydrostatic equilibrium based on $T_{\rm disk, 0} (R)$,
\begin{equation}\label{eq:rho-ini}
	\rho(R,z) = \rhomini (R) \exp\left[  \frac{R^2}{\Hini^2} \pr{\frac{R}{r} - 1} \right]  , 
\end{equation}
where $\rhomini = \Sigma_0 /( \sqrt{2 \pi} \Hini )\propto R^{-q_\rho}$ with $q_\rho$ being $9/4$.
We give the initial poloidal magnetic field from $\bm{B} = \nabla \times (A_\phi \hat{\phi})$ with the toroidal potential vector $A_\phi$
\citep[][as in \citetalias{Bai2017Global-Simulati}]{Zanni2007MHD-simulations},
\begin{equation}
	A_\phi = \frac{2 B_{\rm mid, 0} r_{\text{in}}}{4 - q_\rho - q_T} \left( \frac{R}{r_{\text{in}}} \right)^{-\frac{q_\rho + q_T}{2} + 1} \left[ 1 + (m \tan \theta)^{-2} \right]^{-\frac{5}{8}},
\end{equation}
where the initial vertical magnetic field $B_{\rm mid, 0}$ is given so that a plasma beta $\beta_0$ at the midplane is uniform at $10^4$, and we set $m = 1.0$.

We further set a density floor given by $10^{-10} \pr{r/\rin}^{-q_\rho} \times \rhomini(\rin) $.
We also set a temperature floor given by $50 \,\K\, \pr{r/\rin}^{-q_T}$.
Furthermore, to avoid impractically small time steps from high Alfv\'en speeds in the polar regions, 
we increase the density such that the plasma beta remains above $0.5 (r_{\rm in}/r)^2$ within $R \lesssim \rin$.

%
%


We generally follow the boundary conditions in \citetalias{Bai2017Global-Simulati}. 
The outer boundary conditions are based on an outflow prescription, while the polar boundary conditions follow a reflection model. For the inner radial boundary, we set the poloidal velocities, $v_r$ and $v_\theta$, to zero. 
As a modification from \citetalias{Bai2017Global-Simulati}, 
we give the toroidal velocity by extrapolating the deviation velocity from the Keplerian rotation $(v_\phi - \sqrt{GM_*/r} ) \propto r^{0.5 - q_T}$ inward, 
where we expect the radial balance between the pressure-gradient, gravitational and centrifugal forces.


In regions near the inner boundary, the net poloidal magnetic field threading the disk is artificially truncated, potentially causing artificial features.
We set a buffer region within $r = 2 \rin$ to mitigate the influence from the inner boundary. 
In the buffer region, the diffusivity of the Hall effect is damped in the form of $0.5 + 0.5 \tanh[5 (r - 2 \rin)]$.
Also, the density below $|z| < 4 \Hini$ is relaxed to the initial density structure with 10 local orbital periods.
This approach enables us to conduct long-term calculations and achieve quasi-steady states over a broad range of radii.

Table \ref{tab:parameters} lists all our fiducial simulation parameters, most of which are fixed. We conduct two sets of three simulations, with the main varying parameters listed in Table \ref{tab:model_comparison}. Each set has three simulations with one having no Hall effect, and the other two including the Hall effect but with different field polarities. We fiducially fix the three opacities to be constant as $\kappa_{\rm xuv} =$ 33 \cmcmg, $\kapirr = 1 $ \cmcmg, and $\kappa_{\rm disk} = 1 $ \cmcmg, but we also change $\kapirr$ to 10 \cmcmg to examine its influence on the temperature structure in the second set of simulations.
   






%
%

\subsection{Main Diagnostics}\label{ssec:diag}


In this section, we describe the main diagnostics used in this paper. 
Radial profiles of vertically integrated variables are analyzed with common procedures.
The integration domain is taken to be within $|z| < 5 \Hmid(R)$, where we define it as the disk region 
and $\Hmid(R) = \cs(\Tmid)/\Omega$ is the midplane gas scale height with $\Tmid $ being the midplane temperature.
We denote the vertical integration as,
\begin{equation}
	\int_{\rm disk} ... ~ r \dth .
\end{equation}
The subtraction between the top and bottom values of the disk region is denoted as,
\begin{equation}
	\Big[~  ... ~\Big]^{\rm bot}_{\rm top} .
\end{equation}
Quasi-steady quantities are averaged over time $t \in [300, 500]$ yr, denoted as $\avt{ ... }$. 
In displaying mass and energy change rates, we further smooth them with a Gaussian of width $0.1 r$ to reduce noise.

\subsubsection{Elsasser numbers}
The nonideal MHD effects are characterized by Elsasser numbers.
The Elsasser numbers due to the Ohmic diffusion, Hall effect, and ambipolar diffusion are, respectively, defined by
\begin{equation}
	{\rm Oh} = \frac{ v_{\rm A}^2 }{ \etaO \Omega} ,~
    {\rm Ha} = \frac{v_{\rm A}^2 }{ \etaH \Omega} ,~
 	{\rm Am} = \frac{v_{\rm A}^2 }{ \etaA \Omega} .
\end{equation}
When the Elsasser numbers are smaller than unity, the nonideal MHD effects operate with faster timescales than magnetic induction.
In this paper, we use \Am as a representative indicator of the nonideal MHD effects because \Am is generally independent of the magnetic field strength.



\subsubsection{Mass accretion/loss rate}

The mass accretion rate is a fundamental quantity that
governs disk structure and evolution.
The accretion rate is given by
\begin{equation}
	\dot{M}_{\rm acc}(r) =  -\int_{\rm disk}  2 \pi  R \rho v_r \,r \dth .
\end{equation}
The accretion rate can be compared with that predicted from the equation of angular momentum transport in the disk \citep{Aoyama2023Three-dimension}.
The time-averaged accretion rate is expressed as (see Appendix~\ref{app:acc-rate} for the derivation; \citealt{Aoyama2023Three-dimension})
\begin{equation}
\label{eq:mdot-pred} 
	\avt{ \dot{M}_{\rm acc} (r)} = 
	\dfrac{1}{ \partial j_0 / \partial r } 
	 \pr{ \dfrac{ \partial \avt{J_{r\phi} }}{\partial r}   + \avt{ \gamma_{\theta \phi} }  } ,
\end{equation}
where
\begin{eqnarray}
	\label{eq:j0}
    j_0 &=& \avt{ \left. \int_{\rm disk} 2 \pi R^2 v_\phi r \dth \middle / \int_{\rm disk} 2 \pi R r\dth \right. }, \\
    \label{eq:Jrphi}
    J_{r\phi} &=&  \int_{\rm disk} 2 \pi R^2 T_{r\phi} r\dth , \\
    \label{eq:gamma}
    \gamma_{\theta \phi} &=&  \subtr{ 2 \pi R^2 T_{\theta \phi} } ,
\end{eqnarray}
$j_0$ is the mean specific angular momentum, 
$J_{r\phi}$ is the angular momentum flux due to the $r\phi$-component of the stress tensor, 
and $\gamma_{\theta \phi}$ is the torque density due to the $\phi \theta$-component of the stress tensor.
The $r\phi$-component and $\theta  \phi$-component of the stress tensor are, respectively, defined as,
\begin{eqnarray}
    T_{r\phi} &=&  \rho v_r \delta j / R  - B_r B_\phi/4 \pi ,  \\
    T_{\theta \phi} &=& \rho  v_\theta \delta j / R  -  B_\theta B_\phi/4 \pi  ,
\end{eqnarray}
where $\delta j = R v_\phi - j_0$ is the deviation of the specific angular momentum from the mean value.

In addition, we measure the mass loss rate, which is an indicator of the wind strength.
We define $\dot{M}_{\rm loss}(r)$ as the cumulative wind mass loss rate within radius $r$. 
The derivative form of the mass loss rate is used to measure the local mass loss rate as, 
\begin{equation}
\label{eq:mloss-loc}
	\frac{\d \dot{M}_{\rm loss}(r)}{\d\log r} = 2 \pi R \subtr{ \rho v_{\theta} }. 
\end{equation}
Note that this ``mass loss rate'' can be negative when the gas flows into the disk.



\subsubsection{Energetics}
\label{sssec:energetics}

We evaluate the energy budget by estimating contributions in the total energy equation.
In Appendix~\ref{app:ene-bug}, we derive the approximate energy balance for a quasi-steady state in spherical coordinates, as done for cylindrical coordinates in \citet{Suzuki2016Evolution-of-pr}.
The energy removal rate is due to the disk wind ($L_{\rm wind}$) and radiative cooling ($\Lambda_{\rm rad}$), while the production rate is due to the release of accretion energy ($\Gamma_{r \phi}+\Gamma_{\theta \phi}$) and radiative heating ($\Gamma_{\rm rad}$):
\begin{eqnarray}
 	\label{eq:tot-energy}
     \avt{L_{\rm wind}} + \avt{\Lambda_{\rm rad}} \approx \avt{\Gamma_{\rm acc}} + \avt{\Gamma_{\rm rad}} .
\end{eqnarray}
The wind energy removal rate $\Lwind$ is given by,
\begin{eqnarray}
\label{eq:Lwind}
	L_{\rm wind} &=& 
    \subtr{ \sint \left\{ F_\theta - \rho v_\theta \pr{ \frac{v_0^2}{2} + \Phi} \right\} }  \nonumber \\
	 && - \int_{\rm disk}  \frac{ \cos\theta \rho v_\theta j_0^2 }{R^3} \cdot \sint r \d \theta  , \\
    F_\theta &=&  v_\theta \pr{ E + \rho \Phi + P^*} - \frac{(\bm{v}\cdot\bm{B}) B_\theta}{4 \pi} + S_\theta .
\end{eqnarray} 
The energy production rate $\Gamma_{\rm acc}$ is calculated from the stresses,
\begin{equation}
\begin{split}
    \Gamma_{\rm acc} &= - \int_{\rm disk}  R T_{r\phi} \pd{}{r} \pr{ \frac{v_0}{R}  } \cdot  \sint r \d \theta  \\
                     &  + \subtr{  \sint v_0 T_{\theta \phi}  }. \label{eq:eprod-acc}
\end{split}
\end{equation}
The radiative energy gain rate $\Gamma_{\rm rad}$ and loss rate $\Lambda_{\rm rad}$
are obtained by integrating Equation (\ref{eq:Qrad}), respectively,
\begin{equation}
    \begin{split}
	\Gamma_{\rm rad}  &= \int_{\rm disk} q_{\rm rad, h}  \sint r\dth , \label{eq:GLrad} \\
	\Lambda_{\rm rad}  &= \int_{\rm disk} q_{\rm rad, c}  \sint  r\dth , 
    \end{split}
\end{equation}
where 
\begin{eqnarray}
	q_{\rm rad,h}   &=& \exp\pr{-\taudisk} q_{\rm h,thin}  + \max\pr{ q_{\rm  thick}, 0} ,\\
	q_{\rm rad,c}   &=& \exp\pr{-\taudisk} q_{\rm c,thin}  + \max\pr{ -q_{\rm  thick}, 0} ,\\
	q_{\rm thick} &=& 
		\frac{1}{r^2} \frac{\partial}{\partial r}\pr{r^2 k_{\rm diff} \rho \pd{\cs^2}{r} } .
%
\end{eqnarray}
We here vanish the term on the $\theta$ derivative for the optically-thick term because the defined disk height is higher than the altitude where $k_{\rm diff}$ is cut off ($\taudisk = 1$).




Furthermore, using the internal energy equation, we obtain the energy balance in the mechanical energy (Appendix~\ref{app:ene-bug}). The approximate internal energy balance is given by
\begin{eqnarray}
\label{eq:int-ene-bal} 
    \avt{\Delta \Lambda_{\rm rad}} &+& \avt{L_{\rm wind, int}} 
        \approx \avt{Q_{\rm comp}} + \avt{\Gamma_{\rm Joule}} \\
        L_{\rm wind, int} &=& \subtr{\sint \rho v_\theta e},\\
        Q_{\rm comp} &=& -\int_{\rm disk} \frac{p}{R}\pd{(\sint v_\theta)}{\theta}  r\dth.\\
        \Gamma_{\rm Joule} &=& \int_{\rm disk} \qJ \,r\dth, \label{eq:Gjoule}
\end{eqnarray}
where
$\Delta \Lambda_{\rm rad} = \Lambda_{\rm rad} - \Gamma_{\rm rad}$ is the net radiative cooling rate,
$L_{\rm wind, int}$ is the internal energy loss rate due to the wind,
$Q_{\rm comp}$ is the approximate compressional heating rate,
and $\Gamma_{\rm Joule}$ is the Joule heating rate.
The local Joule heating rate $\qJ$ is given by,
\begin{eqnarray}
	\qJ = \etaO \bm{J}^2 + \etaA \bm{J_\perp}^2 . \label{eq:qjoule} 
\end{eqnarray}
Combining the energy balance of the internal energy with Equation (\ref{eq:tot-energy}), we obtain the energy balance in the mechanical energy:
\begin{eqnarray}
\label{eq:mech-ene-bal}
	\avt{L_{\rm wind,mech}} + \avt{\Gamma_{\rm Joule}}  \approx \avt{\Gamma_{r \phi}} + \avt{\Gamma_{\theta \phi}} - \avt{Q_{\rm comp}} ,
\end{eqnarray}
where $L_{\rm wind,mech} = L_{\rm wind} - L_{\rm wind, int}$ is the mechanical energy loss rate due to the wind.

Finally, to calculate the rate of Joule heating, 
we decompose the current density into
\begin{equation}
	\bm{J} = - \pd{B_\phi}{z}  \hat{\bm{R}}
		  + \pr{ \pd{B_R}{z} - \pd{B_z}{R} } \hat{\bm{\phi}}
		  + \frac{1}{R}\frac{\partial}{\partial R} (R B_\phi) \hat{\bm{z}}.
\end{equation}
The terms with vertical derivatives have been considered in local MHD simulations, while the terms containing $R$ derivatives are due to global field structures. In our later discussion, we further assess the role of the latter contribution, and explicitly write the current density $\bm{J}_{\rm global}$ due to the global structures here:
\begin{equation}
	\bm{J}_{\rm global} = 
		   - \pd{B_z}{R}  \hat{\bm{\phi}}
		  + \pr{ \frac{\partial B_\phi}{\partial R} +  \frac{ B_\phi}{ R} } \hat{\bm{z}}.
\end{equation}
\label{sssec:current}

\subsubsection{Expected Temperature Distributions}
\label{sssec:exp-temp-dist}

To derive the contribution of each heating source to the temperature, we calculate the expected temperature for individual heat sources.
In the analysis, we calculate the temperature profiles expected in the thermal equilibrium for a given heating rate profile $q$ (\citealt[hereafter \citetalias{Hubeny1990Vertical-struct}]{Hubeny1990Vertical-struct}; \citealt[see their Section 2.4.2 for details]{Mori2019Temperature-Str}).
Although the original formula was derived for the vertical direction, we apply it to the $\theta$ direction as a proxy, justified by the disk's cold and thin nature. 
The result is
\begin{eqnarray}\label{eq:texp}
T_{\rm exp} &=& \left [
	 \frac{\mathcal{F}_{\rm top} }{\sigma} 
	\pr{  \frac{3}{4} \tau_{\theta,\rm eff} + \frac{\sqrt{3}}{4} + \frac{q}{4 \rho \kapd  \mathcal{F}_{\rm top}}} \right]^{1/4} ,\\
%
\mathcal{F}_{\rm top} &=& \dfrac{
 \dfrac{\Gamma}{\sqrt{3}}+ {\displaystyle \int_{\theta_{\rm s}}^{\theta_{\rm e}} \rho\kapd \pr{ \int_{\theta_{\rm s}}^{\theta} q \,r \dth' } r\dth }
}{ 2/\sqrt{3} + \tau_{\rm disk, tot} } , \\
\tau_{\theta, \rm eff} &=&  \int_{\theta_{\rm s}}^{\theta} \rho \kapd \pr{ 1 - \frac{1}{\mathcal{F}_{\rm top}} \int_{\theta_{\rm s}}^{\theta'} q \, r\dth'' } \, r\dth'  , \\
\Gamma &=&  \int_{\theta_{\rm s}}^{\theta_{\rm e}} q \,r \dth	, \\
\tau_{\rm disk, tot} &=&  \int_{\theta_{\rm s}}^{\theta_{\rm e}} \rho \kapd \,r \dth , 	
\end{eqnarray}
where 
the $\theta$ boundaries of the analysis domain are $(\theta_{\rm s}, \theta_{\rm e})= (\pi/6, 5 \pi/6)$,
$\mathcal{F}_{\rm top}$ is the re-emitted radiative energy flux at $\theta = \theta_{\rm s}$,
$\tau_{\rm disk, tot}$ is the optical depth over the whole $\theta$ domain,
$\Gamma$ is the integrated heating rate,
and $\tau_{\theta, \rm eff} $ is the effective optical depth contributing to the blanketing effect and depends on $q$.
For symmetric structures, $\mathcal{F}_{\rm top}$ is just half of the heating rate integrated over the full domain.
The temperature near the disk region is insensitive to the choice of the analysis $\theta$-domain. 
Calculating this formula for every radius, we obtain the analytical temperature distribution in the full domain. 

To separately assess the contributions of the different heating mechanisms in the simulation,
we calculate the temperature distributions solely due to the Joule heating, $T_{\rm Joule}$, and irradiation heating, $T_{\rm irr}$.
For the Joule heating, $T_{\rm Joule}$ is given by Equation (\ref{eq:texp}) with the Joule heating rate $\qJ$ (Equation (\ref{eq:qjoule})):
\begin{eqnarray}
	T_{\rm Joule} = T_{\rm exp}(q=\qJ) ,
\end{eqnarray}
where the density distribution and column densities in the simulation are used.
For the irradiation heating, $T_{\rm irr}$ is given by using the irradiation heating rate $\qirr$,
\begin{eqnarray}\label{eq:qirr}
	T_{\rm irr} &=& T_{\rm exp}(q=\qirr) , \\	
	\qirr &=& 4 \rho \kapirr \sigma T_{\rm irr, eq}^4 \exp\pr{- \tauirr}. \label{eq:qirr}
\end{eqnarray}
We have confirmed that 
the radial temperature profile calculated for an irradiated disk model agrees with the result of a Monte Carlo simulation (see Appendix~\ref{app:anal}).

To further assess the efficiency of the Joule heating as opposed to viscous heating,
we also calculate the temperature distribution solely due to the viscous heating. 
We assume a viscosity (taken to be constant) that yields an accretion rate that is the same as found in the simulation. 
We can calculate the radial profile of the midplane temperature as
 \begin{eqnarray}
	\Tvis &\approx& \left [ \frac{  \Gamma_{\rm vis} }{2 \sigma} \pr{ \frac{\kappa_{\rm disk} \Sigma}{4}  + \frac{1}{\sqrt{3}} } \right]^{1/4} , \label{eq:Tvis} \\
	\Gamma_{\rm vis} &=&  \frac{  3 \dot{M} \Omega^2 }{4 \pi} ,
\end{eqnarray}
where $\Gamma_{\rm vis}$ is the energy production rate in the viscous disk,
and $\kappa_{\rm disk} \Sigma / 4$ is the effective optical depth for a viscous heating rate distribution proportional to the gas density (see Equation (A8) in \citealt{Mori2021Evolution-of-th}).

\subsubsection{Static model}
\label{sssec:static-model}
To emphasize changes in the temperature distribution induced by disk dynamics,
we calculate the temperature of irradiated disks in hydrostatic equilibrium for comparison.
We refer to this model as the Static model.
This model is independent of the simulation results and is solely determined by the surface density distribution and opacities.

The model calculation is done by the following iterative process.
First, given our fiducial surface density profile, we calculate the density structure assuming a prior midplane temperature profile and that the disk is vertically isothermal.
The temperature distribution in thermal equilibrium is then calculated by Equation (\ref{eq:texp}) for the whole domain, where the heating rate is given by $\qirr$ (Equation (\ref{eq:qirr})).
Next, we fit the obtained midplane temperature profile outside $r >$ 0.8 au\footnote{
We find that temperature profiles outside $r \gtrsim 1 $ au can be well fitted with a power-law profile, while the temperature profile inside is affected by the inner boundary, and thus is neglected here.
} with a power-law dependence on a radius, and then pass the fitting information to the next iterative cycle (again assuming the disk is vertically isothermal).
This fitting procedure usually stabilizes after a few iterations, and the fitted midplane temperature profile is generally in good agreement with the temperature profile from the radiative transfer calculation.
We then define the resulting fitted midplane temperature profile as the Static model, given by
$156.3 \,\K ~(R/\au)^{-0.469}$ for $\kapirr = 1 $ \cmcmg and $165.5 \,\K ~(R/\au)^{-0.458}$ for $\kapirr = 10 $ \cmcmg.
We consider this profile to be more accurate than
the two-layer model for irradiation heating \citep{Chiang1997Spectral-Energy} (to be further discussed in Section~\ref{ssec:comp-model} and Appendix~\ref{app:anal}), as it incorporates more self-consistent treatment of the irradiation heating.

\section{Simulation Results} \label{sec:results}

\begin{deluxetable}{lccc}
\tablecaption{Simulation Models \label{tab:model_comparison}}
\tablehead{
\colhead{Name} &  \colhead{Field polarity} & \colhead{$\kapirr$ [\cmcmg]} 
}
\startdata
\rFNH &  No Hall & 1  \\
\rFHp &  $+$ & 1  \\
\rFHn &  $-$ & 1   \\
\rONH & No Hall & 10  \\
\rOHp &  $+$ & 10 \\
\rOHn &  $-$ & 10  
\enddata
\end{deluxetable}

We start by discussing the simulation without the Hall effect (\rFNH; Section \ref{ssec:res-nohall}), followed by those with the Hall effect:
with aligned field polarity (\rFHp; Section \ref{ssec:res-hall+}) and anti-aligned field polarity (\rFHn; Section \ref{ssec:res-hall-}).
Additionally, we discuss cases with larger stellar irradiation opacity (\rONH, \rOHp, \rOHn; Section \ref{ssec:opt-opac}). 
Then, in Section~\ref{ssec:hall-irr-acc}, we discuss the novel phenomenon of episodic accretion.

\subsection{Case without Hall Effect} \label{ssec:res-nohall}

\begin{figure*}
    \centering
   \includegraphics[width=\linewidth]{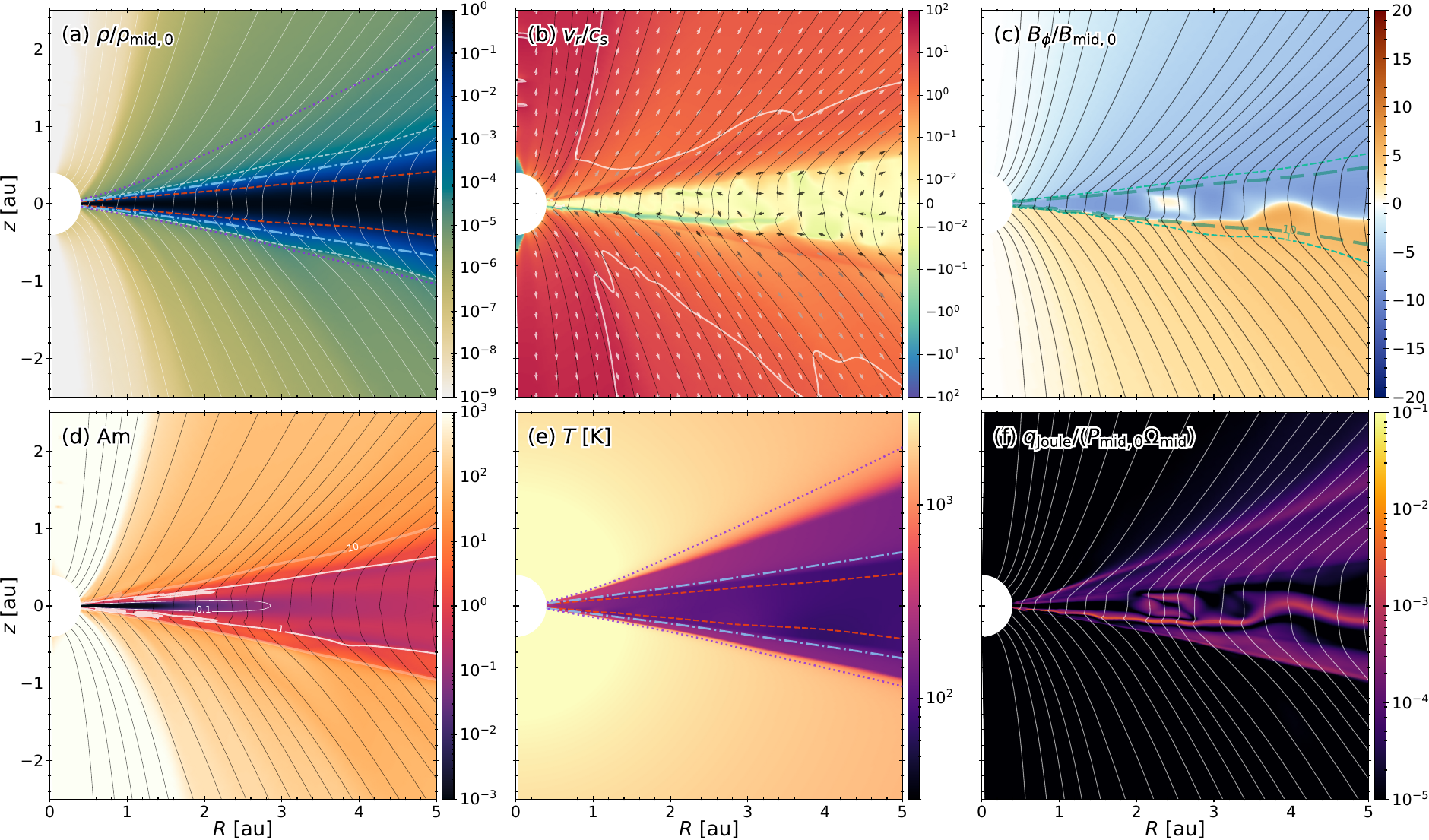}
    \caption{
    Two-dimensional dynamical structures of the inner region at $t = 365$ yr, in \rFNH run.
    (a): 
    gas density normalized by the initial midplane density with the surfaces of $\tauxuv = 1$ (purple dotted line), $\tauirr = 1$ (blue dotted-dashed), $\taudisk = 1$ (red dashed), and the disk region $z = 5 \Hmid$ (white dashed).
    Thin lines threading to the disk show the poloidal magnetic field.  
    (b): 
    Mach number of the radial gas velocity, with the poloidal field line.  
    The white solid line represents the location where the poloidal speed $|\bm{v}_{\rm pol}|$ 
    reaches the poloidal Alfv\'en velocity (Alfv\'en surface).
    The arrows show the poloidal velocity direction (black: subsonic, white: supersonic).
    (c): 
    toroidal magnetic field normalized by the initial magnetic field strength at the midplane, with the poloidal field line.
    Green contour lines correspond to the plasma beta $\beta$ = 1 (thin), and 10 (thick).
    (d): 
    ambipolar Elsasser number ${\rm Am}$, with the poloidal field line. 
    The white contours show the level of 0.1 (dashed), 1 (solid), and 10 (dotted).
    (e): 
    gas temperature, with the same optical depth and height lines as in panel (a). 
    The snow surface ($T = 170$ K) is shown in the cyan line.
    (f): Joule heating rate, with the poloidal field lines.
    \label{fig:noH-map}
    }
\end{figure*}

In this subsection, we briefly discuss the general properties of the fiducial Hall-free simulation. Given its similarity to previous studies, we then primarily focus on a more detailed discussion on the disk temperature structure.

\subsubsection{General properties}


\begin{figure*}
\centering
   \includegraphics[width=0.86\linewidth, clip]{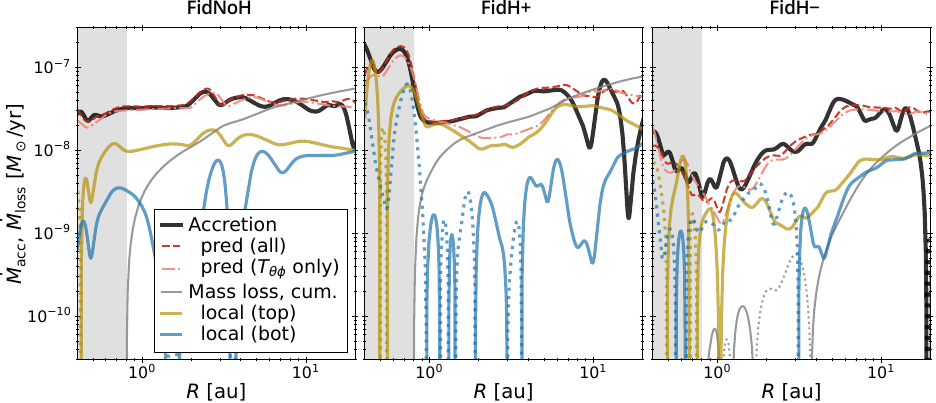}
    \caption{        
        Radial distributions of the time-averaged mass-accretion rate ($\avt{\dot{M}_{\rm acc}}$; black thick) and cumulative mass-loss rate ($\dot{M}_{\rm loss}$; gray thin) for the runs of \rFNH (left), \rFHp (middle), and \rFHn (right).
        The accretion rates predicted from the wind stress $T_{\phi \theta}$ only (orange dash-dotted) and from both $T_{r \phi}$ and $T_{\phi \theta}$ (red dashed; see Equation (\ref{eq:mdot-pred})) are shown.
        The local mass loss rates ($\d \dot{M}_{\rm loss} / \d \ln r $; Equation (\ref{eq:mloss-loc})) of the top and bottom surfaces are, respectively, shown with blue and yellow lines, with negative values indicated by dotted lines.
        The accretion rate and local mass-loss rate are smoothed. 
        The shaded represents the buffer zone (see Section~\ref{ssec:simulation-setup}).
    \label{fig:Mdot-dists}
    }
\end{figure*}

\begin{figure*}
\centering
      \includegraphics[width=0.86\linewidth, clip]{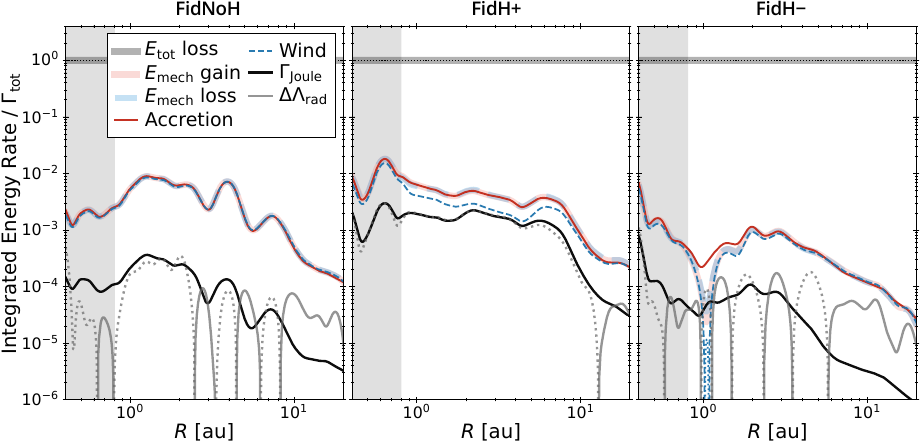}   
    \caption{
       Radial distributions of the time-averaged energy change rates normalized by the energy gain in the total energy equation 
       ($\Gamma_{\rm acc} + \Gamma_{\rm rad}$; Equations (\ref{eq:tot-energy})) with smoothing:
       the loss of the total energy ($\Lambda_{\rm wind} + \Lambda_{\rm rad} $; gray thick; Equation (\ref{eq:tot-energy}))),
       loss of mechanical energy ($\Lambda_{\rm wind, mech} + \Gamma_{\rm Joule}$; dashed light blue; Equation (\ref{eq:mech-ene-bal})),
       gain of mechanical energy ($\Gamma_{\rm acc} - Q_{\rm comp}$; red; Equation (\ref{eq:mech-ene-bal})),
       release by the disk accretion ($\Gamma_{\rm acc}$; yellow; Equation (\ref{eq:eprod-acc})),
       loss by the disk wind ($\Lambda_{\rm wind}$; dashed dark blue; Equation (\ref{eq:Lwind})),
       dissipation by Joule heating ($\Gamma_{\rm Joule}$; black; Equation (\ref{eq:Gjoule})),
       and net radiative cooling ($\Lambda_{\rm rad} - \Gamma_{\rm rad} $; gray thin; Equation (\ref{eq:GLrad})).  
    \label{fig:efrac-dists}
    }
\end{figure*}

Figure \ref{fig:noH-map} shows the two-dimensional distributions of the gas density, radial velocity, toroidal magnetic field, Elsasser number of ambipolar diffusion, temperature, and volumetric heating rate at $t = 365$ yr (165 yr after the activation of radiation transport).
The magnetic field evolves and reaches a steady state in $\sim$ 10 local orbital time, as mentioned in \citetalias{Bai2017Global-Simulati}.
This snapshot represents typical physical structures in the steady state.
The basic structures of the density and magnetic field are consistent 
with previous global nonideal MHD simulations \citep{Gressel2015Global-Simulati, Gressel2020Global-Hydromag, Bai2017Global-Simulati, Iwasaki2024Dynamics-Near-t}, which we discuss only briefly in this section.
Nevertheless, we note that the disk aspect ratio, $h = \Hmid/R$, is smaller than in the previous global MHD simulations because the disk temperature is lower (see the next section).

Disk winds are launched from both the top and bottom surfaces of the disk (panels (a) and (b) in Figure \ref{fig:noH-map}), as in the previous studies. While it is often found that wind launching occurs at the FUV ionization front \citepalias[\citealt{Bai2016Magneto-thermal};][]{Bai2017Global-Simulati} where the gas approaches the ideal MHD regime, we note that in our case, the field lines already start to bend outwards lifting the gas flows in a region where $\Am \sim 1$--$10$, below the FUV front (panel (d) in Figure \ref{fig:noH-map}):
This is related to the fact that the FUV front in our simulation is located high in the atmosphere, and in this case, wind launching can take place in regions where gas is not fully coupled to magnetic field with $\Am\gtrsim1$.

The toroidal magnetic field is developed within the disk which must flip across the disk to conform to the favorable geometry for wind launching. With very low level of disk ionization to support current at the inner disk, this flip must occur at one side of the disk surface (\citealp{Bai2013aWind-driven-Acc}; panel (c) in Figure \ref{fig:noH-map}), 
which is a general phenomenon \citep{Wang2024Nonideal-Magnet}.
At larger distances ($\gtrsim$ 3--5 au), the flip can take place closer to the midplane, consistent with \citetalias{Bai2017Global-Simulati}.

The top/bottom surfaces show significant differences in the mass loss rates, as shown in Figure \ref{fig:Mdot-dists}, where the local mass loss rates differ by 4--20 times within 5 au. Such asymmetric winds have also been observed in earlier studies, such as \citet{Riols2020Ring-formation-}. The difference is associated with the strong asymmetry in the magnetic field configuration discussed earlier. We also see that the difference in local mass loss rates becomes less pronounced towards the outer disk region, where the asymmetry is also weaker.
 The accretion rate is largely constant at $\sim$3$\times 10^{-8} \Msun/\yr$ (Figure \ref{fig:Mdot-dists}), which is well consistent with the prediction from the total stresses (Equation \ref{eq:mdot-pred}).
It is also consistent with the accretion rate predicted by the wind stress alone, indicating that the accretion is primarily wind-driven.

To further understand the basic energetics, we plot the primary energy change rates in the total energy equation (Equation (\ref{eq:tot-energy}); \citealt{Suzuki2016Evolution-of-pr}), shown in Figure \ref{fig:efrac-dists}.
The change rates are normalized by the gain of the total energy, which is the sum of the radiative heating term and the energy release term due to the gas accretion.
As the loss of the total energy (the radiative cooling term and wind loss) is normalized to unity, we see that the total energy is balanced (the residual is $<$ 0.1\%). 
Additionally, since the gain and loss of $E_{\rm mech}$ coincide perfectly, we see that the mechanical energy is also balanced (see Equation (\ref{eq:mech-ene-bal})).
The accretion energy release rate is approximately equal to the wind loss $L_{\rm wind}$,
while the Joule dissipation is a minor contributor.
Furthermore, the gas compressibility and the internal energy loss due to the wind are negligible in the overall energy balance.






\subsubsection[]{Temperature Structure}
\label{ssec:nohall-temp}

\begin{figure*}
    \centering
   \includegraphics[width=0.98\linewidth, clip]{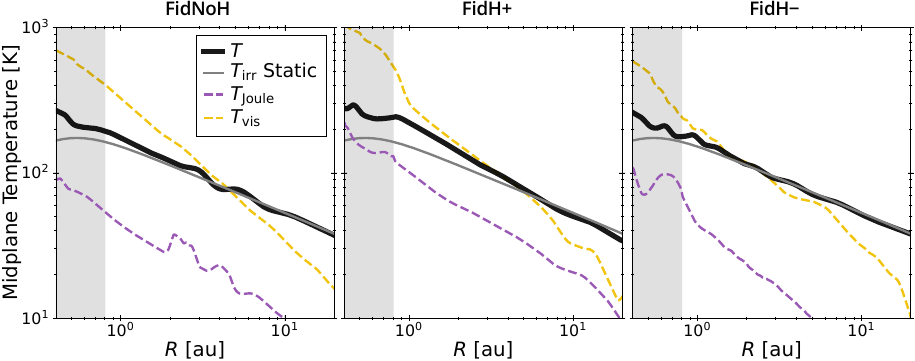}
    \caption{
   Radial distributions of the but for the time-averaged midplane temperature:
   the simulation temperature (black), expected temperature solely due to the Joule heating rate that is calculated from the simulation ($T_{\rm Joule}$; purple dashed), the viscous disk model ($\Tvis$; Equation (\ref{eq:Tvis}); orange dashed), and the Static model (gray solid; see Section~\ref{sssec:static-model}).   
    \label{fig:temp-profs}
    }
\end{figure*}


In this subsection, we focus on the disk temperature structure as shown in panel (e) in Figure \ref{fig:noH-map}. It can be divided into three regions: the XUV heated (within a column of $\Sigma_r < 0.03$ \gcmcm based on our prescription, purple dotted line), (optical-)irradiation heated (immediately below till blue dash-dotted line), and re-emission-heated regions (in between the blue dash-dotted lines). We have verified that the simulation temperature structure matches well with that from the Monte Carlo simulation (see Appendix~\ref{app:test}). 
Among the three regions, the gas temperature in the XUV and irradiated regions reaches the equilibrium temperature assumed in Equations (\ref{eq:Tirr})--(\ref{eq:Txuv}). Given that our treatment of the XUV heating is highly simplified and that this region is largely optically thin, it has little effect on the overall disk temperature structure. We therefore primarily focus on the bulk disk regions heated by stellar irradiation and re-emission.


We first examine the midplane temperature profile in Figure \ref{fig:temp-profs}, and compare the results with different models. 
First, we have confirmed that the midplane temperature is thermally relaxed within $\sim 30$ yrs.
We also see that the midplane temperature in the simulation is close to that in the Static model, especially in the outer region (Figure \ref{fig:temp-profs}). 
When assuming disk temperature is solely determined by the Joule heating, the predicted temperature would be significantly lower. 
This suggests that the overall disk temperature is primarily determined by the irradiation heating. 
On the other hand, when assuming viscous heating at the given wind-driven accretion rates, one would obtain a steep temperature profile with higher temperatures at the inner disk, which always well exceeds that resulting from the Joule heating.
Note that earlier global disk simulations usually prescribe relatively large disk thickness to enhance resolution (e.g., \citetalias{Bai2017Global-Simulati} assumed the disk aspect ratio to be $h = 0.045$ at $r = 1$ au), 
whereas with more realistic radiation transport, we find the midplane temperature to be $\approx$ 180 K at $r = 1$ au, leading to $h=0.027$.

\begin{figure*}
    \centering

    \includegraphics[width=0.88\linewidth]{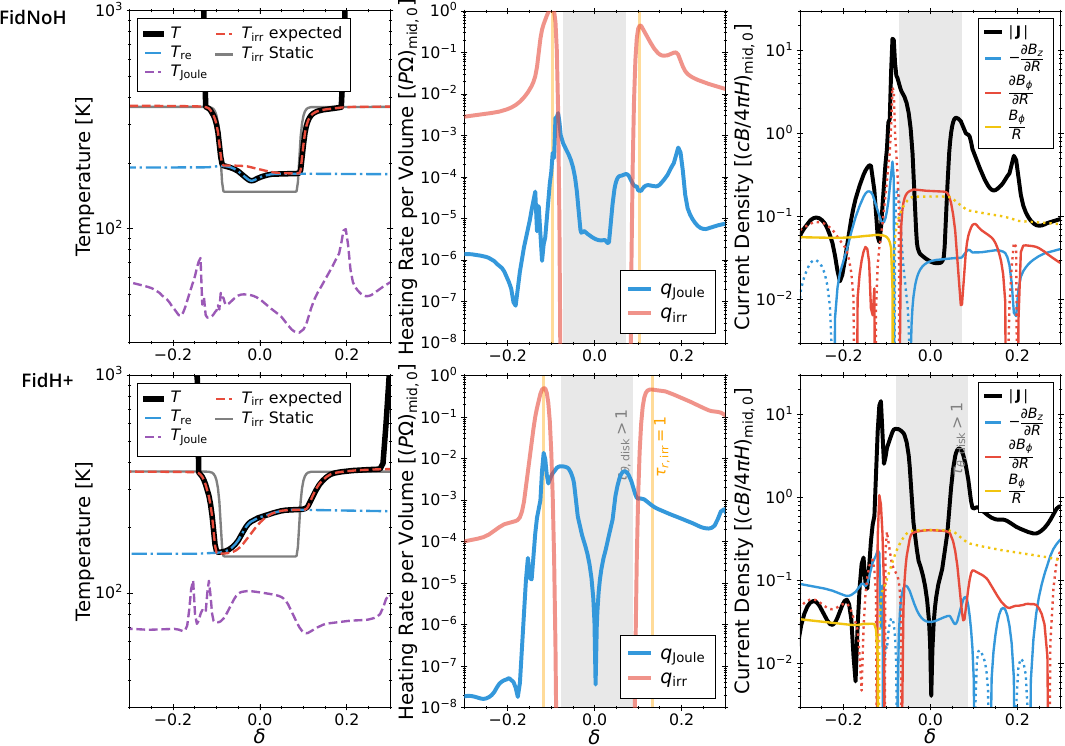} 
    \caption{
    Vertical profiles of the temperature (left), heating rate per unit volume (middle), and current density (right) at $r = 1$ au and $t = 365$ yr, for the \rFNH (top panels) and \rFHp (bottom panels).
       Left: simulation gas temperature (black solid; among which the contribution from re-emission is shown by blue dash-dotted), and expected temperatures due to Joule heating only (purple dashed), due to irradiation only under the Static model (gray solid), and under the simulation density and heating rate profiles (red dashed).
    Middle: Joule heating rate ($\qJ$; blue) and irradiation heating rate ($\qirr$; orange) are shown.
    Gray shaded region is optically thick for disk radiation. 
    Vertical yellow lines indicate the location of $\tauirr = 1$.
	Right: magnitude of the current density (black) and the components due to global field structure (colored; see Section~\ref{sssec:current}), where negative values are shown by dotted lines. 
    \label{fig:noH-qT} \label{fig:Hp-qT}
    }
\end{figure*}


%
We next analyze the vertical temperature profile and associated heating mechanisms at 1 au in Figure \ref{fig:noH-qT}. 
From the top left panel, we first confirm that the temperature profile well agrees with that expected from pure irradiation heating, with the midplane temperature
well reproduced from the re-emission heating.
Interestingly, we find that the expected irradiation temperature is slightly asymmetric. This arises from the difference in the rate by absorbed irradiation energy between the top and bottom surfaces, caused by asymmetric outflows (see Section~\ref{sssec:hall+-temp}). 
When the disk is optically thick, re-emission from each surface largely independently determines the midplane temperature on each side.
Although the asymmetry in the temperature profile is relatively small in this run, we see a larger asymmetry in the \rFHp run (Section~\ref{sssec:hall+-temp}).

In the top middle panel of Figure \ref{fig:noH-qT}, we analyze the contribution from different heating mechanisms.
We have seen that accretion heating due to magnetic diffusion (i.e., Joule heating) is highly inefficient. 
Besides that Joule heating is inherently weak, it is also more concentrated in the disk upper layer where the horizontal fields flip (around $\Dth \sim -0.06\text{--}0.08$ at $r = 1$ au).
This has been seen in earlier
local MHD simulations \citep{Mori2019Temperature-Str,Bethune2020Electric-heatin}, and it has been known that surface heating is inefficient in warming up the disk midplane region.
We also show the magnitude of the current density $|\bm{J}|$ and the terms due to the global effects in the upper right panel in Figure \ref{fig:noH-qT} (see Section~\ref{sssec:current}).
We see that the current density from global gradients is almost always much smaller than the total current density. This is related to the fact that we find $B_\phi \propto R^{-1}$ in the bulk disk, making the $\partial B_{\phi}/\partial R$ and $B_\phi/R$ terms largely cancel out. Therefore, the current from the global gradients are generally negligible.

\subsection{Case with the Hall Effect: Positive Polarity} \label{ssec:res-hall+}

We next present the results of the \rFHp run, which includes the Hall effect with the positive field polarity.
As in Section~\ref{ssec:res-nohall}, we first analyze the physical and temperature structures at $t = 365$ yr in Sections \ref{sssec:hall+-ss}--\ref{sssec:hall+-temp}.

\subsubsection{General properties}
  \label{sssec:hall+-ss}
  
\begin{figure*}
    \centering
   \includegraphics[width=\linewidth]{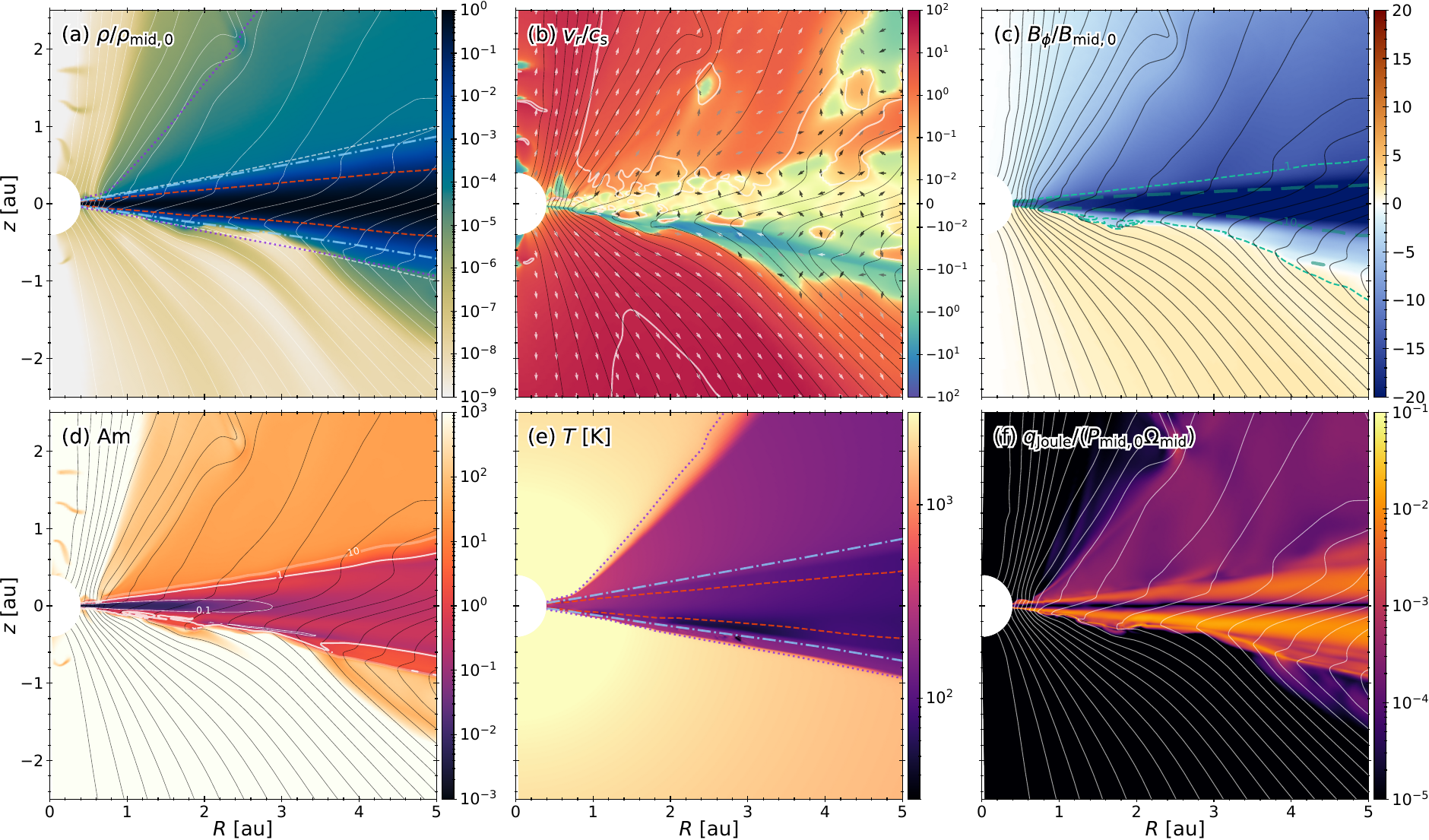}
    \caption{
    Same as Figure \ref{fig:noH-map}, but for \rFHp. 
    \label{fig:Hp-map}
    }
\end{figure*}


%
%
%

As in Figure \ref{fig:noH-map}, we first show the two-dimensional distributions of the basic physical variables at $t = 365$ yr in Figure \ref{fig:Hp-map}.
In the disk region, the radial and toroidal magnetic fields get strongly amplified due to the HSI \citepalias[\citealt{Kunz2008On-the-linear-s};][]{Bai2017Global-Simulati}.
The amplified field pushes out the current layer to one side of the disk surface where the horizontal field flips \citepalias[e.g.,][]{Bai2017Global-Simulati},
while within a substantial height of the disk $|\Dth| \lesssim 0.2$, the direction of the horizontal field remains the same.
Most of the wind-driven accretion goes through the surface current layer, where the horizontal field flips and exerts most of the stress. 
We find that the accretion velocity in the layer can reach up to $\approx 10 \cs$, which is $\approx$ 30\% of the Keplerian velocity. 
On the other hand, due to the HSI that amplifies both $B_r$ and $B_\phi$ that substantially enhances the radial Maxwell stress,
the contributions to net accretion rate by $\Trp$ and $\Tpt$ are comparable beyond $R = $ 2 au (Figure \ref{fig:Mdot-dists}), as in \citetalias{Bai2017Global-Simulati}.

Intriguingly, the simulation clearly shows a strong asymmetry in the disk wind.
While the asymmetry in the wind density is seen in the \rFNH run, the density contrast in the \rFHp run is significantly higher.
For example, at $R \approx 4$ au, the density at $z = + 1$ au is three orders of magnitude higher than that at $z = - 1$ au.
Note this strongly asymmetric disk wind occurs after the activation of the Hall effect, and the low-density side is the side where the current sheet is located.

\begin{figure}
    \centering
        \includegraphics[width=\linewidth]{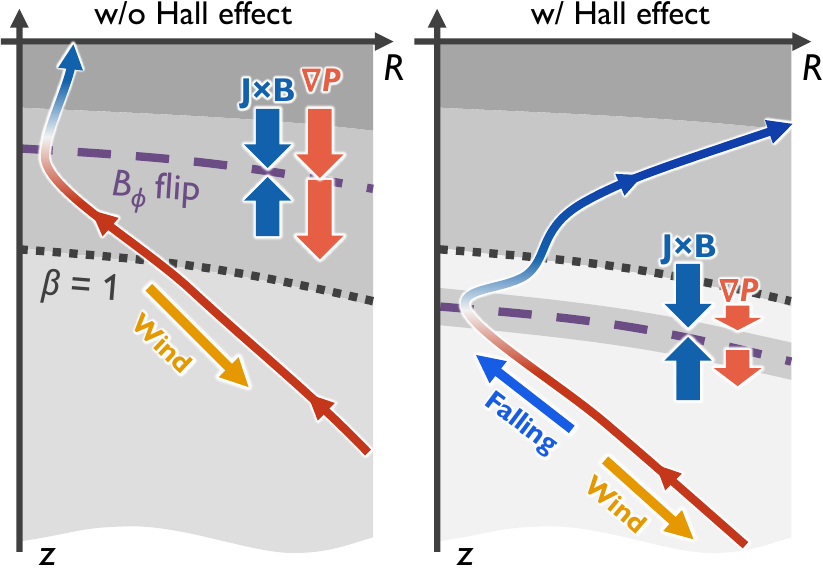}
    \caption{
Schematic illustration of the mechanism for forming a low-density wind when the Hall effect is active especially for the positive polarity. Left: Without the Hall effect, the current layer, where $B_\phi $ reverses, is located in the gas-pressure-dominated region, and the wind is typically launched above this layer near where $\beta \sim 1$. 
Right: With the Hall effect, the current layer is pushed to a higher altitude than the $\beta \sim 1$ surface. The Lorentz force dominates over the pressure gradient force and draw the gas into the current layer, creating a downward flow in the surface.
As a result, the wind base is shifted to even higher altitudes, thus with much lower density.
\label{fig:low-dens-wind}
    }
\end{figure}

By inspecting the vertical structure, we identify the following mechanism likely responsible for strongly reducing the wind density on the side with the current sheet, as illustrated in Figure \ref{fig:low-dens-wind}.
In the run \rFNH, wind launching empirically occurs around the region where the magnetic pressure dominates over the thermal pressure (i.e., plasma $\beta \sim 1$). The current sheet where $B_\phi$ flips resides below the $\beta\sim1$ layer. In run \rFHp, with additional magnetic field amplification thanks to the HSI, the current sheet is elevated compared to the Hall-free case, reaching the $\beta\sim1$ region or above. In this case, the Lorentz force primarily acts to drive the accretion flow, with an additional vertical component (from $J_rB_\phi$, again dominates over gas pressure) that
draws the gas toward the current layer, leading to a downward flow.
Wind launching thus occurs even at a higher location.
Therefore, with both the elevation of the current sheet, together with the downward flow above the current sheet, the gas density at the wind base is substantially reduced, and hence the wind density itself.
We also note that while the downward flow tends to drag magnetic flux with it, this is balanced by ambipolar diffusion that
prevents excessive magnetic flux buildup above the disk, thereby sustaining the general field configuration.
There is a significant difference in the mass loss rate from the two sides of the disk, as shown in Figure \ref{fig:Mdot-dists}.
The local mass loss rate $\d M_{\rm loss}/\d \ln r$ on the top surface is only slightly higher than that in \rFNH run,
but local mass loss rate from the bottom side of the surface is smaller by about one order of magnitude.
Especially, there are radii where the ``mass loss rate'' is negative, which corresponds to downward flow (see Equation (\ref{eq:mloss-loc}) for the definition).
In fact, the system is never steady and there are cycles of 
gas lifted from the outer region and then fall back to an inner region (see Section \ref{ssec:hall-irr-acc}).
Regarding the energy balances as seen from Figure \ref{fig:efrac-dists},
the overall result is similar to the \rFNH run.
The main difference is that while the disk wind dominates energy loss, which balances the release of the accretion energy, the Joule dissipation plays a more significant role due to strong amplification of the magnetic field by the HSI.
The Joule heating rate accounts for $\sim$ 20--50 \% of the accretion energy rate, which is consistent with results from the local MHD simulations \citep{Mori2019Temperature-Str}. 

\subsubsection{Temperature structures}\label{sssec:hall+-temp}

Panel (e) of Figure \ref{fig:Hp-map} shows the two-dimensional temperature distribution. 
First, we discuss the temperature in the wind region.
With the asymmetric winds, the temperature structure becomes notably asymmetric as well.
In the atmosphere on the upper side, the wind shields the XUV radiation, and thus the thermal irradiation largely determines the temperature, as in \rFNH.
The lower side is mostly heated by the XUV due to the lower column density, leading to a thinner irradiation region.
On the lower side, gas clumps in the accretion layer (e.g., at $R \sim $1--2 au) cast shadows on the outer surface with lower temperatures, which will be further discussed in the next subsection.


Next, we discuss the temperature structure in the disk body and analyze the results in similar ways as in the Hall-free case by examining the middle panel of Figure \ref{fig:temp-profs} as well as Figure \ref{fig:Hp-qT}.
We see that the Joule heating rate is higher compared to \rFNH (see Figure \ref{fig:efrac-dists}).
Nevertheless, the heating occurs at a relatively high altitude of $\sim 2$--3$H$, reducing the effective optical depth significantly ($\tau_{\theta, \rm eff}$ in Equation (\ref{eq:texp})). As a result, the contribution of the accretion heating is still minor compared to the irradiation heating and thus does not strongly influence the temperature structure. 
Although the current density around the midplane is enhanced by the global field gradients to a certain extent (bottom right panel in Figure \ref{fig:Hp-qT}), it is still not sufficient to change the overall picture. 
By contrast, assuming viscously driven accretion to achieve the same accretion rate, the resulting viscous heating is again much stronger than the Joule heating. 




\begin{figure*}
    \centering
    \includegraphics[width=\linewidth, clip]{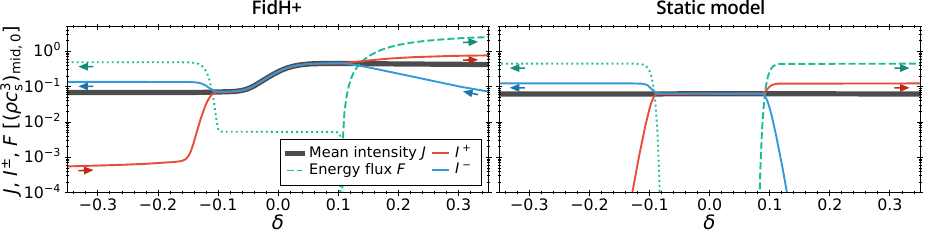}
    \caption{
    Vertical profiles of the mean radiative intensity $J$ (black) and energy flux (green; positive shown with dotted) for the \rFHp run (left) and Static model (right), where negative values are shown with dotted lines.
    The radiative intensities in the upward ($\Dth \to \infty$; $I^+$, red) and downward ($\Dth \to -\infty$; $I^-$, blue) directions are shown.
    Small arrows depict the direction of each radiative variable. 
    \label{fig:Hp-J}
    }
\end{figure*}


Interestingly, although the accretion heating is inefficient, 
the midplane temperature becomes higher than in the passive-disk model (i.e., Static model) and in the \rFNH run (see Figure \ref{fig:temp-profs}), especially within $\sim$ 4 au. 
To a first look, this is related to the fact that the vertical temperature profile shows more significant temperature asymmetry within the bulk disk than the \rFNH run (lower left panel in Figure \ref{fig:Hp-qT}).
Figure \ref{fig:Hp-J} shows the vertical profiles of the mean radiation intensity $J$ and upward/downward radiation intensity $I^{\pm}$ (see Appendix~\ref{app:Tre}) in the simulation and the Static model. 
In the lower side  ($\Dth < 0$), the profiles of $J$, $I^{\pm}$, and energy flux $F$ are similar to the Static model.
In the upper side ($\Dth > 0$), $J$ is larger than the Static model.
This is because the high-density disk wind absorbs more irradiation energy, and thus re-emits stronger radiation toward the midplane ($I^{-}$).
In short, it is the denser disk wind that intercepts more irradiation and leads to stronger irradiation heating.


\subsection{Case with the Hall Effect: Negative Polarity}
\label{ssec:res-hall-}

\begin{figure*}
    \centering
  \includegraphics[width=\linewidth]{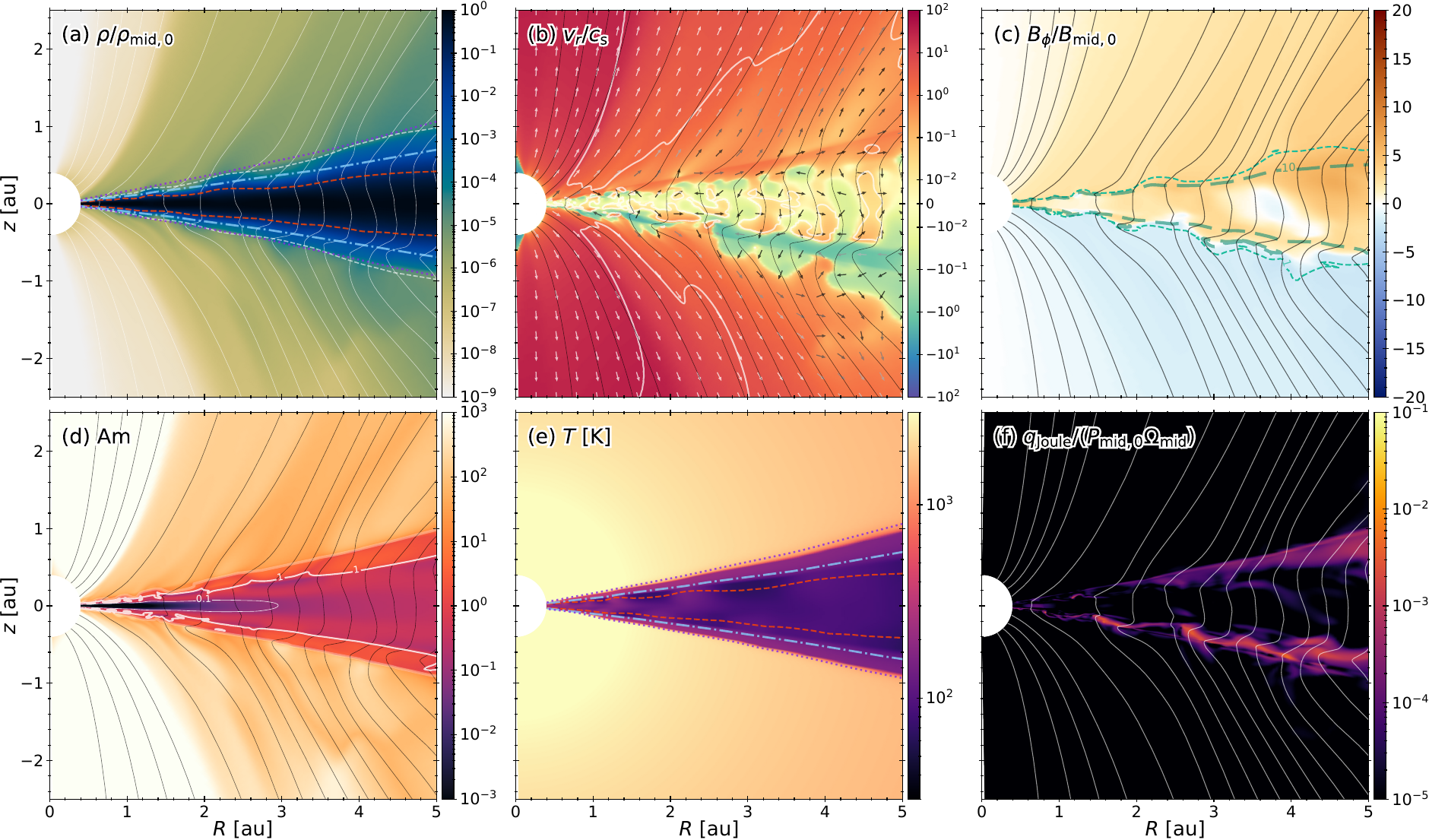}
    \caption{
    Same as Figure \ref{fig:noH-map}, but for \rFHn .
    \label{fig:H--map}
    }
\end{figure*}




This subsection presents the case of the negative field polarity (\rFHn), in comparison to \rFNH and \rFHp (Sections~\ref{ssec:res-nohall} and \ref{ssec:res-hall+}).
The parameters are the same as those in \rFHp, except for the direction of the magnetic field.
In this case, rather than the HSI to amplify the horizontal field, the Hall effect here reduces the horizontal magnetic field.

\subsubsection{General properties}

We first show in Figure \ref{fig:H--map} the two-dimensional structures of the main physical quantities, similar to Figure \ref{fig:noH-map}.
The overall results are similar to the anti-aligned case presented in \cite{Bai2017Global-Simulati}: with the Hall effect reducing horizontal fields, the field lines are largely vertical through the bulk disk.
Nevertheless, this configuration is weakly unstable and the toroidal field of one sign still fills the bulk disk, and flips on one side of the disk. In our simulation, this occurs also in the lower side, leading to asymmetric accretion flows largely concentrating on the lower disk surface.

With a substantially reduced midplane field, angular momentum transport by the $r\phi$ component of the Maxwell stress becomes much weaker compared to the \rFHp run, and accretion is almost entirely wind-driven, similar to the \rFNH case (see Figure \ref{fig:Mdot-dists}). The wind density and the wind mass loss rate are lower compared to run \rFNH, which is related to the reduction of the horizontal field and hence magnetic pressure support in the surface.

In the meantime, the poloidal magnetic flux is rapidly transported outward, as discussed in \citet{Bai2017Hall-Effect-Med}, and the mean poloidal field at $R \sim 1$ au gets weaker than the \rFNH run by 20\% by the end of the simulation.

\subsubsection{Temperature structure}
\label{sssec:hall--temp}

We here show the temperature structure in the \rFHn run and clarify the influence of the field polarity. 
The two-dimensional temperature distribution is shown in the panel (e) in Figure \ref{fig:H--map}.
As the wind density is smaller than the \rFHp run, the stellar optical radiation penetrates and warms up the wind.

To examine how the disk temperature is shaped, we show the time-averaged temperature profile on the midplane in Figure \ref{fig:temp-profs}.
The overall distribution is similar to the temperature profile in the Static model.
We see that the contribution of the accretion heating is negligible, which can be understood from the previous discussions, as follows.
The energy dissipation of the accretion heating occurs far from the midplane, which is optically thin for radiative cooling. 
Additionally, the efficiency of the energy dissipation in the disk is even lower (the right panel in Figure \ref{fig:efrac-dists}).
The energy dissipation fraction in the accretion energy is $\sim 10^{-2}$ --$10^{-1}$.
Furthermore, the enhancement of the irradiation heating by the disk wind is not as efficient as \rFHp because of the smaller wind density.



\begin{figure}
    \centering 
            \includegraphics[width=.99\linewidth, trim={0 0mm 0mm 0mm}, clip]{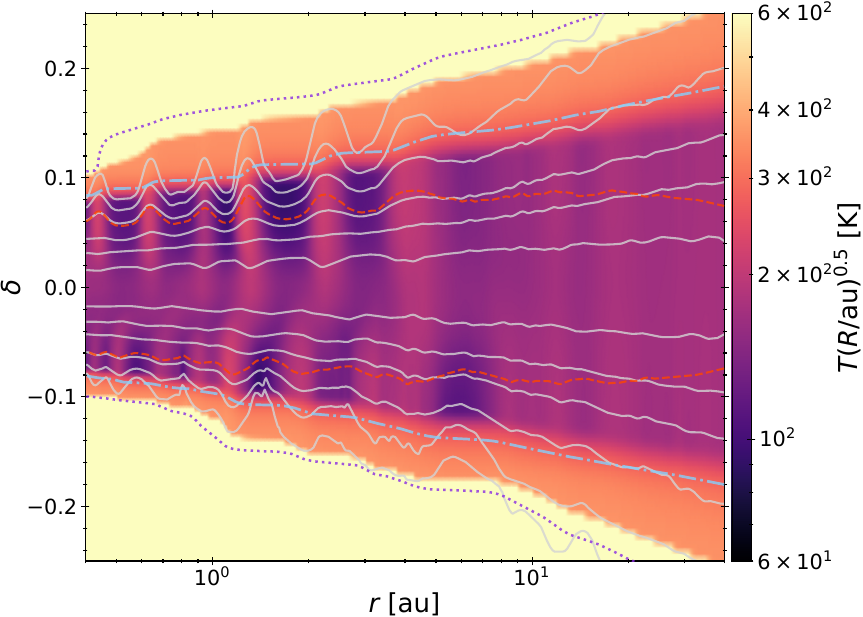}
    \caption{
    Temperature distribution normalized by $(R/\mathrm{au})^{-0.5}$ at $t = 365$ yr in the \rFHn model, with the $x$-axis being the logarithmic radius and $y$-axis being $\delta$. We plot the surfaces where optical depth equals unity  ($\tauxuv = 1$; purple dotted line,  $\tauirr = 1$; blue dotted-dashed line, $\taudisk = 1$; red dashed line), and iso-density contours at levels \{$10^{-4}, 10^{-3}, 10^{-2}, 10^{-1}, 0.3, 0.7$\} from the midplane.
    \label{fig:H-_temp-dens-dist}
    }
\end{figure}

In addition, we found bumpy features in the two-dimensional temperature structure that correspond to the density bumps on the disk surface. To illustrate this correlation, Figure \ref{fig:H-_temp-dens-dist} shows iso-density contours overlaid on the temperature distribution, which is normalized by $(R/\mathrm{au})^{-0.5}$.
This temperature structure is reminiscent of the self-shadowing instability, which is also referred to as a thermal wave or irradiation instability (e.g., \citealt{DAlessio1999On-the-Thermal-,Watanabe2008Thermal-Waves-i,Ueda2021Thermal-Wave-In,Wu2021The-Irradiation,Okuzumi2022A-global-two-la,Kutra2024Irradiated-Disk,Chrenko2024The-Inner-Disk-}; however see \citealt{Melon-Fuksman2022No-Self-shadowi,Pavlyuchenkov2022aSimulation-of-T}).
In this instability, stellar irradiation heats the perturbed surfaces, raising the local disk temperature and hence the local scale height. This makes the disk more flared and then intercepts more stellar irradiation.
\citet[]{Melon-Fuksman2022No-Self-shadowi} argues that the instability does not occur because thermal waves get damped by radiative diffusion before affecting the midplane.
As we do not observe such temperature variations in run \rFNH, we may tentatively assert that our results are consistent with theirs, and attribute the presence of such density and temperature variations to the complex interplay between irradiation, winds, and non-ideal MHD effects. Such variations also leads to time variabilities, and we proceed with further discussion in Section \ref{ssec:hall-irr-acc}.

\subsection{Influences of the Irradiation Opacity} \label{ssec:opt-opac}

\begin{figure*}
 \centering
      \includegraphics[width=0.94\linewidth, clip]{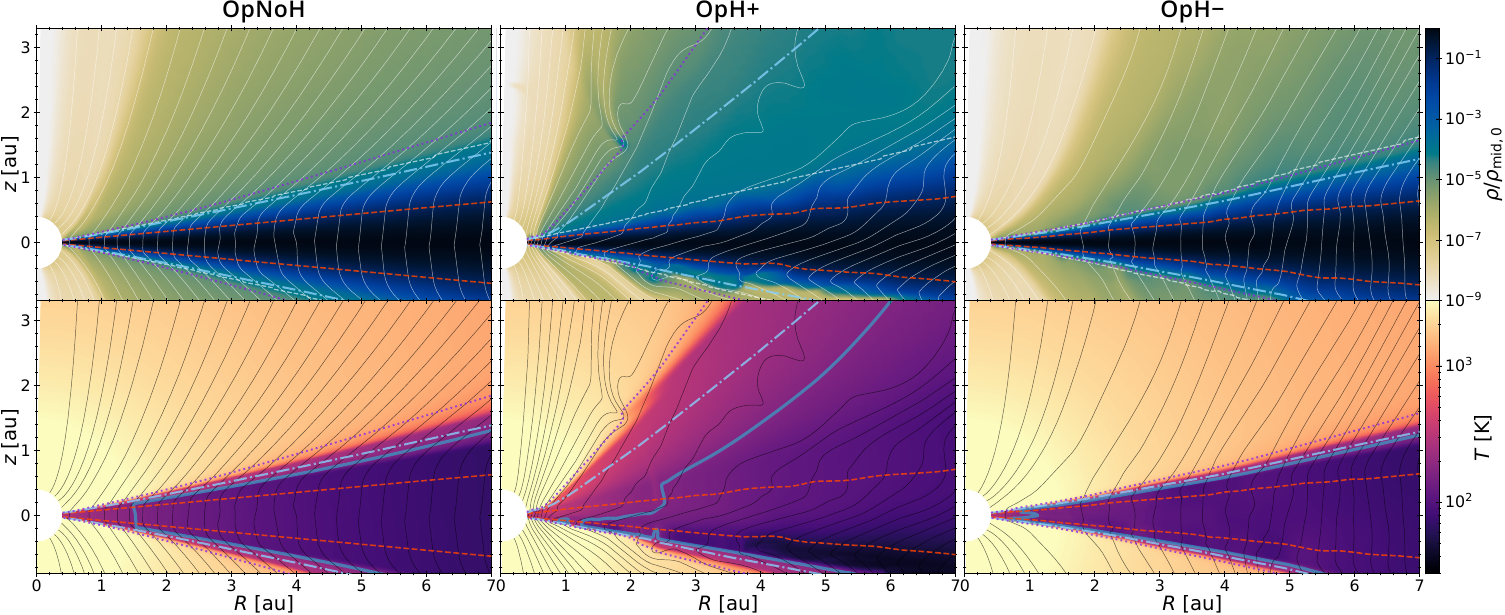}    
    \caption{
        Two-dimensional distributions of the density (upper panels) and temperature (lower panels) for $\kapirr = $ 10 \cmcmg at $t = 365$ yr: \rONH (left), \rOHp (middle), and \rOHn (right).
        The lines are the same as in (a) and (e) of Figure \ref{fig:noH-map}. Note that the irradiation front is shown with cyan dotted-dash lines. 
The upper wind region is mainly displayed.
    \label{fig:op-dists}
    }
\end{figure*}
%
%
%

In Section~\ref{sssec:hall+-temp}, we have proposed that the disk temperature can be enhanced due to the elevated irradiation front by the dense disk wind. Here, we further validate this effect by increasing the optical opacity for stellar irradiation $\kapirr$: we increase $\kapirr$ to 10 \cmcmg from 1 \cmcmg used in previous sections. 
This effectively mimics the situation of dust entrainment in disk winds \citep[e.g.,][]{Rodenkirch2022Dust-entrainmen}.


Figure \ref{fig:op-dists} illustrates how the irradiation fronts are affected by this increase in $\kappa_{\text{irr}}$. 
The density structures of the wind in \rONH and \rOHp are similar to the cases with $\kapirr = 1$ \cmcmg.
The \rOHn run also shows a similar density structure to \rFHn, except for the density fluctuation on the disk surface (discussed below).
The \rOHp run, however, shows that the irradiation front is significantly elevated to $\sim 40$ degrees with respect to the midplane and lies in the wind region, 
as compared to $\sim 9$ degrees in run \rFHp.
For other higher-opacity runs, although the irradiation fronts remain close to the disk surface, they are also elevated to a modest level compared the fiducial cases.

\begin{figure*}
    \centering
   \includegraphics[width=0.98\linewidth]{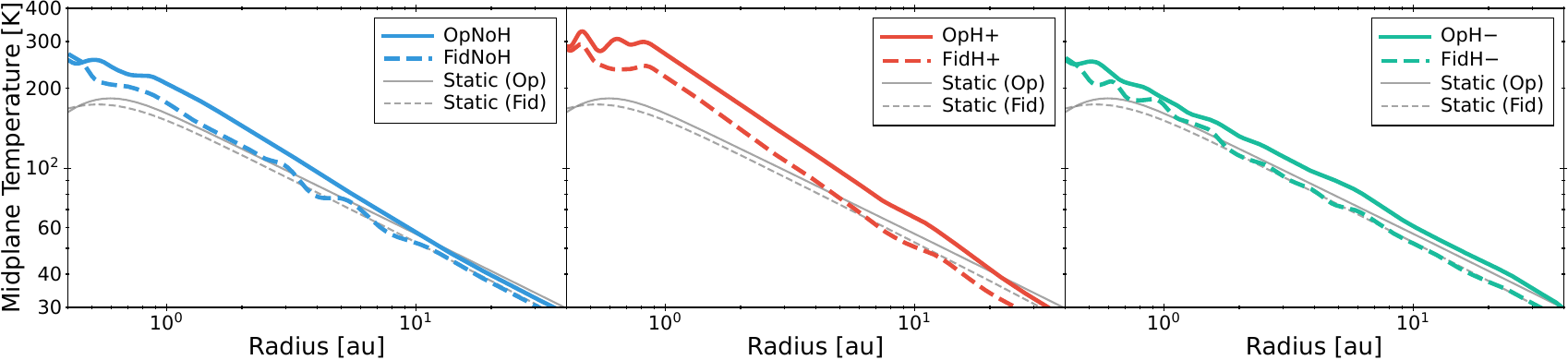}
    \caption{
         Time-averaged midplane temperature profiles with $\kapirr = $ 10 \cmcmg: \rONH (left), \rOHp (middle), and \rOHn (right).
        For comparison, the models with $\kapirr = $ 1 \cmcmg are shown as colored dashed lines, and the Static model as thin gray lines.
    \label{fig:all_temps}
    }
\end{figure*}



Figure \ref{fig:all_temps} shows the midplane temperature profiles in comparison with the lower opacity models. As expected,
our higher opacity simulations consistently show higher midplane temperatures, typically by $\sim20\%$.
By conducting analysis similar to that in Section~\ref{ssec:nohall-temp}, we have confirmed that the accretion heating is still negligible in all the cases. This means that the enhancement of the midplane temperature is primarily due to the enhancement in the irradiation heating.

\begin{figure*}
    \centering
   \includegraphics[width=0.98\linewidth]{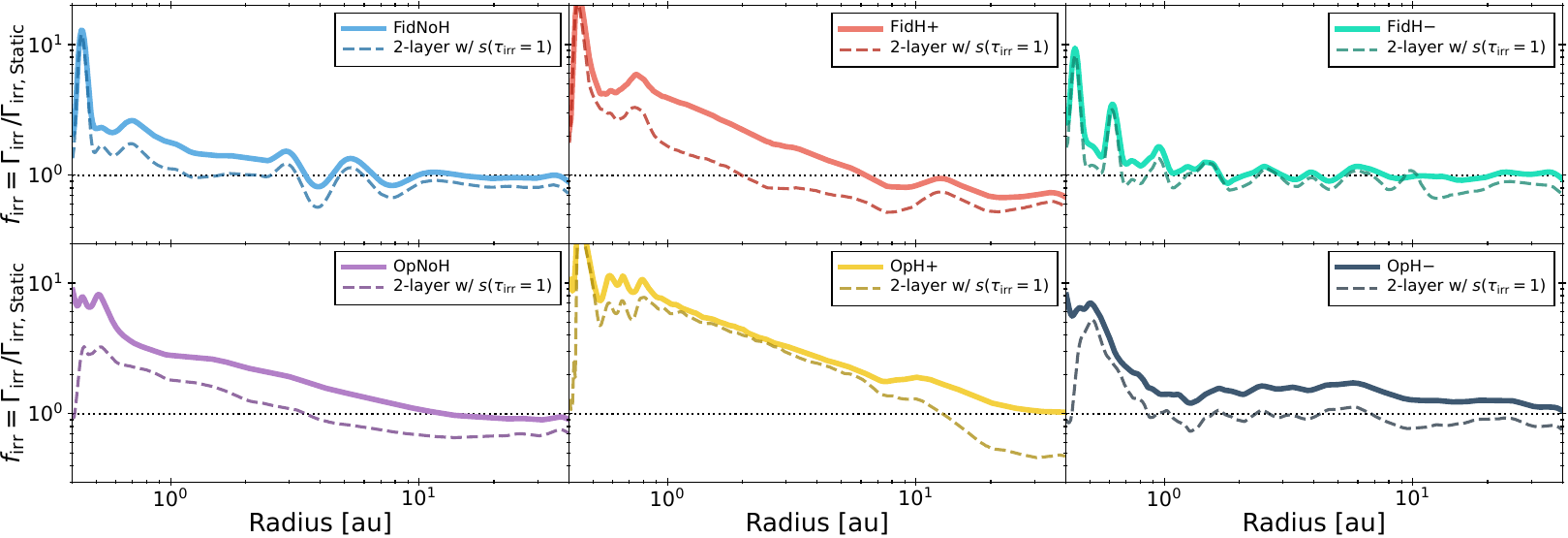}
    \caption{
    Ratio of the total irradiation heating rate in the simulation to that in the Static model, $f_{\rm irr}$, after the time average.
    All the runs in this paper are shown:   
         \rFNH (upper left), \rFHp (upper middle), \rFHn (upper right),  \rONH (lower left), \rOHp (lower middle), and \rOHn (lower right).
         Dashed lines represent the energy-rate ratio based on the two-layer model with the absorption surface of $\tauirr = 1$ (see Section~\ref{ssec:comp-model} and Appendix~\ref{app:anal}).
    \label{fig:firr}
    }
\end{figure*}

To quantitatively assess the enhancement of the irradiation heating, Figure \ref{fig:firr} introduces $f_{\rm irr}$, the ratio of $\Gamma_{\rm irr}$ in our simulation to that in the Static model, where 
\begin{equation}
    \Gamma_{\rm irr} = \int_{\rm disk} q_{\rm h, thin} \sin \theta r \d \theta . 
\end{equation}
This value represents how much the heating rate is enhanced by the wind.
We find that within $\sim 10$ au, heating rate from our simulations is generally higher than in the Static model.
In particular, this effect is more significant in the inner a few au region.
We confirm that the enhanced heating rates correspond to the increase in the midplane temperature from the Static model.
For example, in the most notable case of the \rOHp run, the heating rate at $r \sim$ 1 au is boosted by $\approx$ 7 times, leading to a midplane temperature increase by a factor of $7^{0.25} = 1.68$. 
In summary, these facts support the effect of the wind-enhanced irradiation heating.

We can speculate on the conditions required for wind-enhanced irradiation heating, which should be related to the optical depth of the disk wind.
In the \rFHn run,  very little enhancement of the irradiation heating is observed (see Section~\ref{fig:temp-profs}, or Figure \ref{fig:all_temps}). 
From this, we infer that a mass loss rate higher than that in \rFHn is likely necessary to trigger wind-enhanced irradiation heating. 
Considering the opacity, we simply propose that the condition for wind-enhanced irradiation heating in the inner few au region is approximately 
$	\kapirr \cdot \d M_{\rm loss} / \d \log r \gtrsim 1 \cmcmg \times 3 \cdot 10^{-9} \Mpyr $, assuming wind velocities insensitive to models.


Intriguingly, in addition to the temperature structure, we find that the increase in the opacity changes the behavior of the disk dynamics in the \rOHn run.
As shown in Figure \ref{fig:op-dists}, the density and temperature fluctuations in the \rOHn run are remarkably reduced compared to \rFHn.
In addition, this is also the case in the \rOHp run (Figures \ref{fig:all_temps}).
These facts suggest that the higher $\kapirr$ weakens or suppresses the action of the irradiation instability.
We attribute this phenomenon to the fact that as the irradiation front is lifted due to higher opacity, it essentially enters or at least approaches the wind-launching region, and any perturbations can be quickly advected away by the disk wind and/or the surface accretion flow.

\subsection{Hall-Irradiation-induced variability of surface accretion}
\label{ssec:hall-irr-acc}

One novel phenomenon we observe in runs with the Hall effect is that accretion through the strong current layer is episodic, leading to substantial variability in accretion rates.
As the wind launching region is further up, the wind kinematics also becomes highly variable.
This phenomenon occurs after radiation transport is activated.
In addition, it does not occur in the runs without the Hall effect.
In this section, we describe the time-variable surface accretion and identify a potential mechanism that explains such variabilities.

\subsubsection{Phenomenological overview}

\begin{figure*}[t]
    \centering 
       \includegraphics[width=.9\linewidth, trim={0 8.8mm 0 0}, clip]{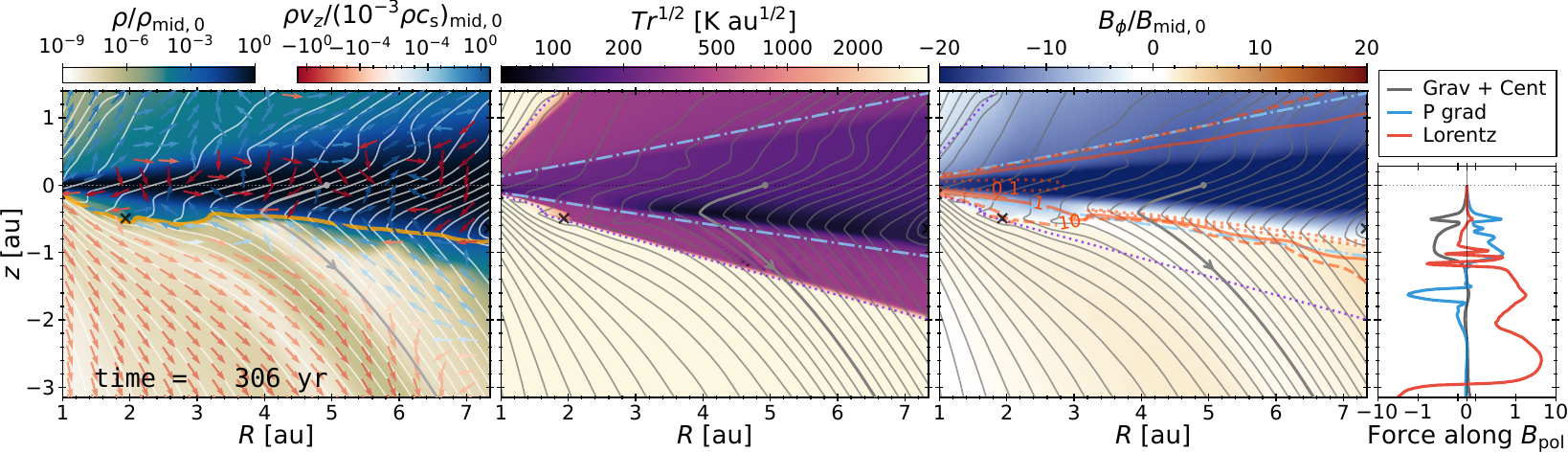}\\
       \includegraphics[width=.9\linewidth, trim={0 8.8mm 0 15mm}, clip]{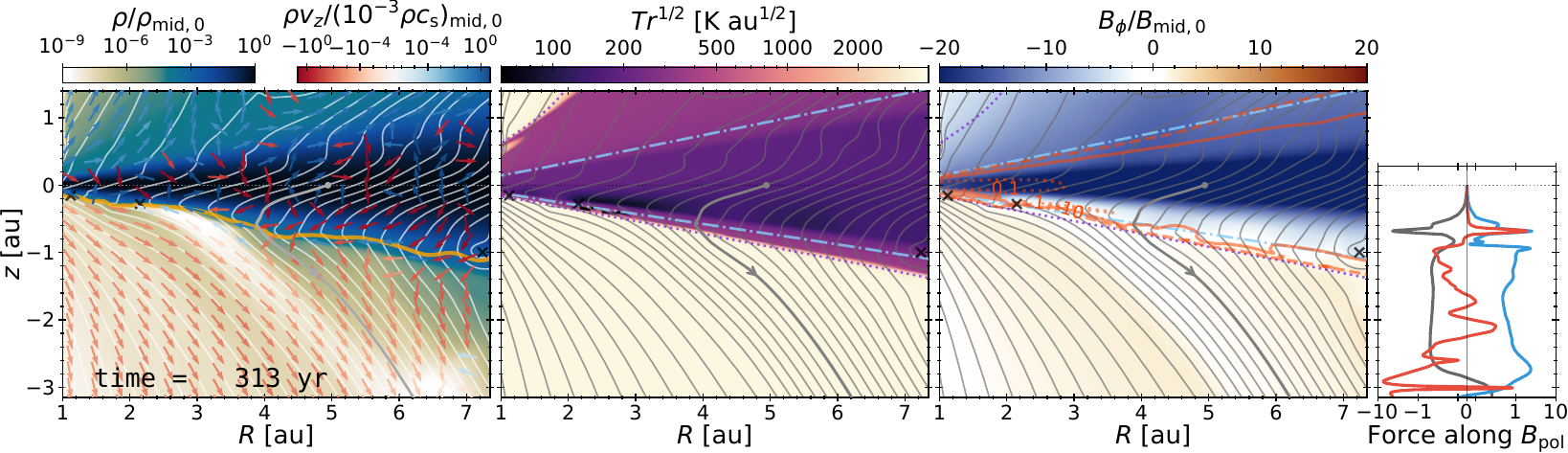}\\
       \includegraphics[width=.9\linewidth, trim={0 8.8mm 0 15mm}, clip]{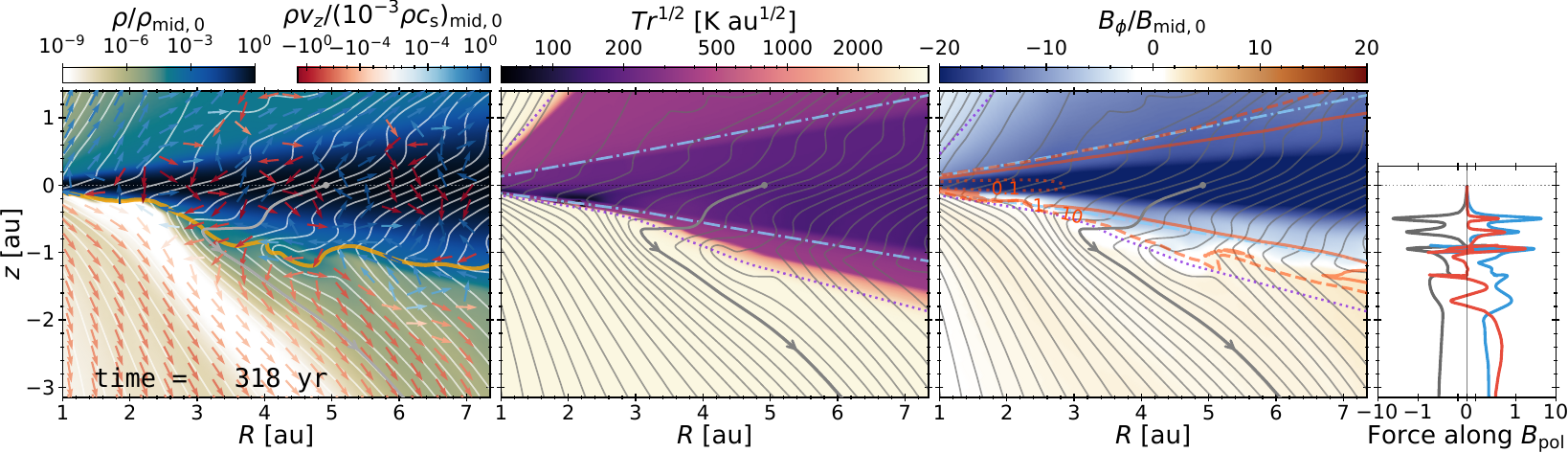}\\
       \includegraphics[width=.9\linewidth, trim={0 0mm 0 15mm}, clip]{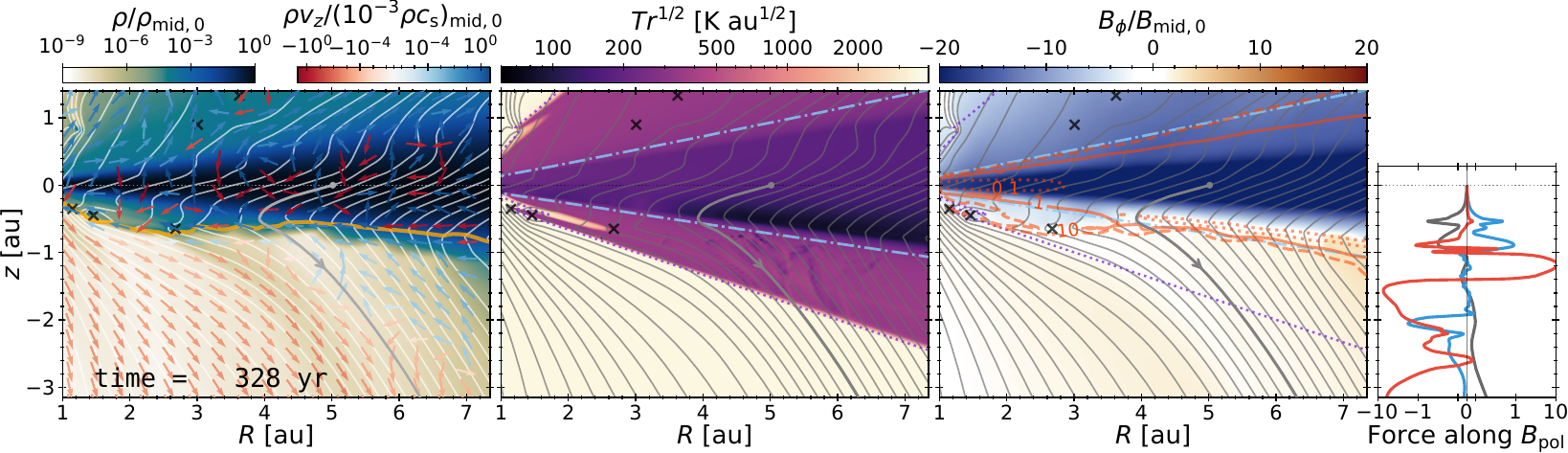}\\
       \vspace{-.6em}
    \caption{
    Snapshots of the physical structures on the southern disk surface at $t =${306, 313, 318, 328} yr (top to bottom rows) in the \rFHp run. 
    Left maps: density with velocity vectors (color: vertical mass flux), and poloidal magnetic fields (thin lines). 
    Orange lines indicate the current layer where the toroidal field flips.
    Crosses mark the locations of magnetic islands, which are the centers of looped poloidal fields.
    Middle maps: temperature scaled by $r^{1/2}$, with lines of $\tauxuv = 1$ (dotted) and $\tauirr = 1$ (dash-dotted). 
    Right maps: toroidal magnetic fields with $\Am$ contours ($\Am = 0.1$ [red dotted], $1$ [red solid], $10$ [red dashed]), and the same lines as in the middle panels. 
    Rightmost panels: forces along the poloidal fields (i.e., a positive force helps launch the outflow)  originating from 5 au (indicated in all other panels). The forces present the sum of gravitational and centrifugal forces (gray), the pressure gradient force (blue), and the Lorentz force (red), each normalized by the gravitational force.
    \label{fig:H+_timevar}
    }
\end{figure*}

Figure \ref{fig:H+_timevar} shows a series of time snapshots of the density, temperature and magnetic fields during one cycle of such repeated events in the \rFHp run.
At $t = 306$ yr, in the inner region ($R \sim 2$ au), there is a gas clump on the surface, from which the wind is launched. 
Meanwhile, the outer region ($R \sim $3--10 au) is shadowed from the inner disk, the gas falls onto the disk surface, and wind-launching occurs at much higher altitudes.
By $t = 313$ yr, the inner clump gets accreted and diminishes, and the outer disk is no longer shadowed. We observe that the wind is launched from the outer disk surface, increasing the gas density in the lower-side disk atmosphere. As the disk surface lifted by the wind accretes toward the inner region, an extended accretion layer is formed ($t \approx 318$ yr). The accretion flow makes poloidal magnetic fields kink inward. Such kinked field lines reconnect and make
this layer fragment into clumps, which become the new inner clumps ($t = 328$ yr), and this leads to the beginning of the next cycle.
Typically, the cycle repeats itself approximately every $\approx$ 30 yrs.

\begin{figure}
    \centering
       \includegraphics[width=.97\linewidth, trim={0 8.0mm 0mm 0.mm}, clip]{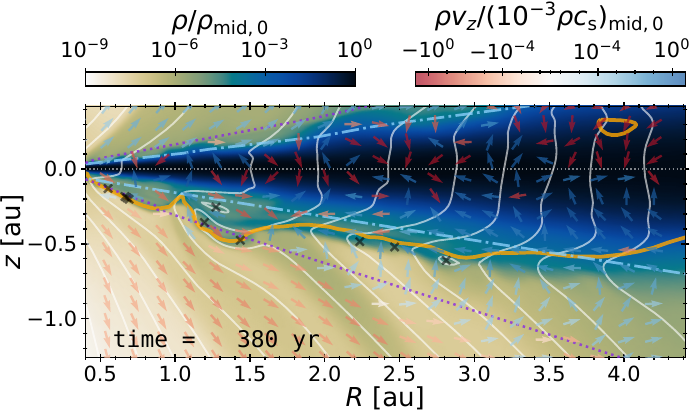}\\
       \includegraphics[width=.97\linewidth, trim={0 8.mm 0mm 17.mm}, clip]{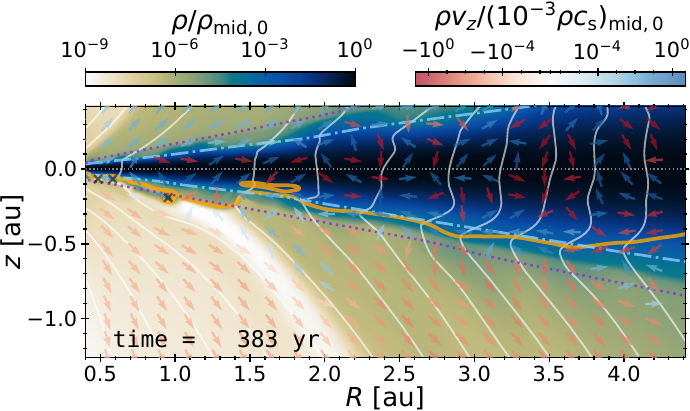}\\
       \includegraphics[width=.97\linewidth, trim={0 8.mm 0mm 17.mm}, clip]{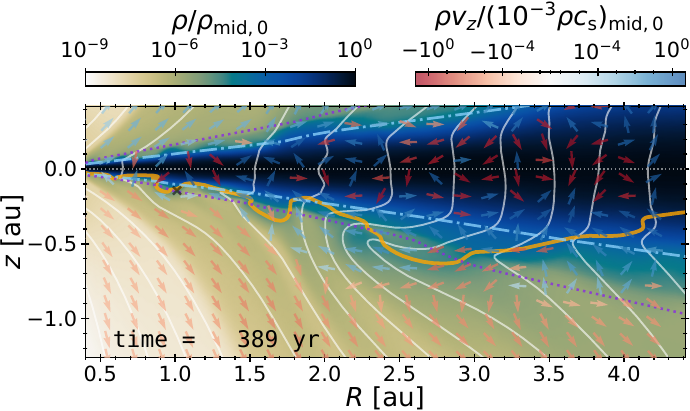}\\
       \includegraphics[width=.97\linewidth, trim={0 0mm 0mm 17.mm}, clip]{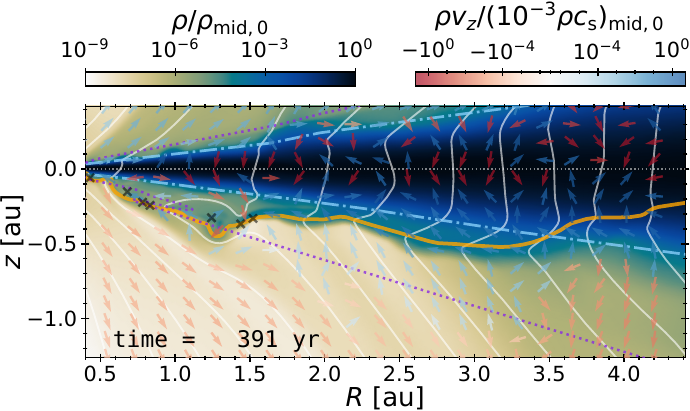}       
    \caption{
    Same as the left panels in Figure \ref{fig:H+_timevar}, but for the \rFHn run with the snapshots at $t =$\{380, 383, 389, 391\} yr. 
        The lines of $\tauxuv = 1$ (dotted) and $\tauirr = 1$ (dash-dotted) are also shown. 
    \label{fig:H-_timevar}
    }
\end{figure}

A similar episodic accretion occurs in the \rFHn run.
Figure \ref{fig:H-_timevar} shows a series of snapshots of the density structure during one cycle.
After the inner clumps shrink during surface accretion, we observe that
a wind is launched around $R \sim 3$ au ($t = 383$ yr).
The lifted gas quickly accretes inward ($t = 389$ yr), forming new inner clumps that cast a shadow on the outer region ($t = 391$ yr).
Compared to the \rFHp run, the accreting clumps are smaller in size.
Also, the observed
``duty cycle" is $\approx$ 5 yr, which is shorter than in the \rFHp run.


\begin{figure*}
 \centering
%
    \includegraphics[width=0.92\linewidth, clip]{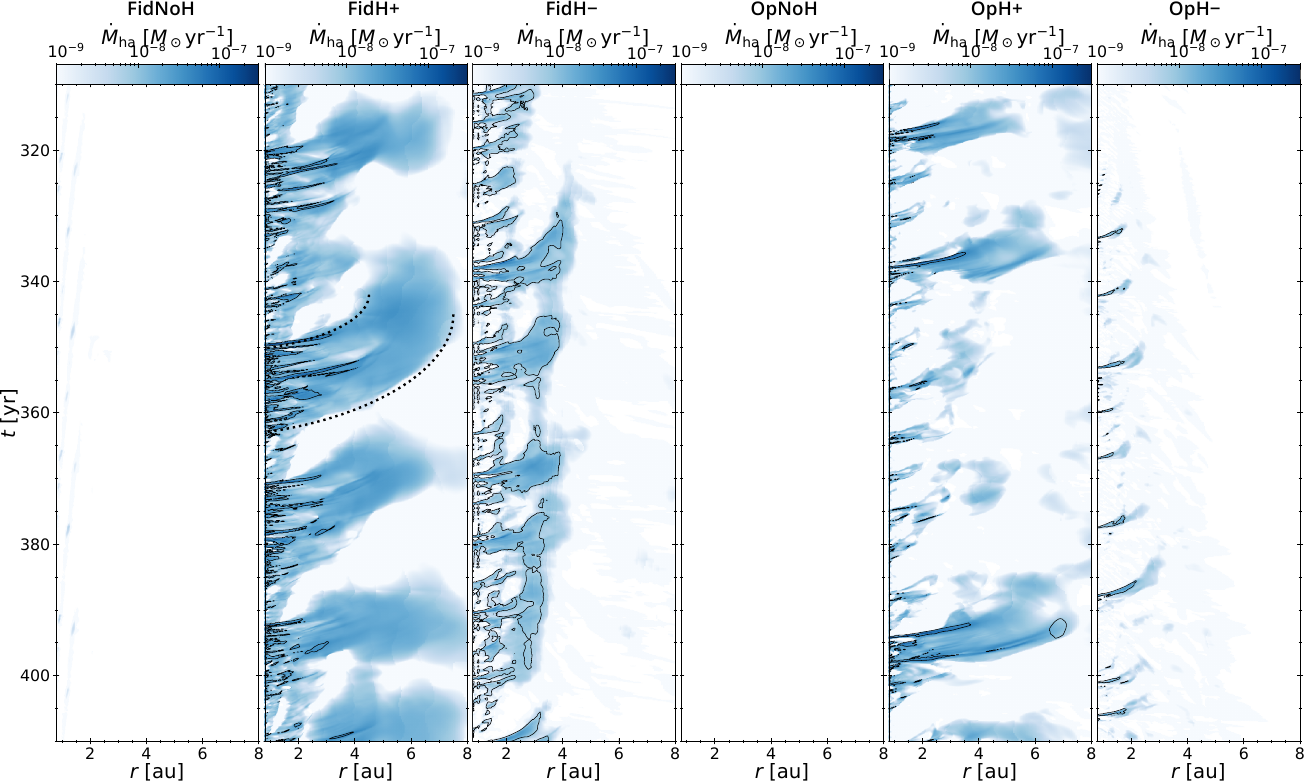}
    \caption{
    Space-time distribution of the accretion rate in the atmospheric region on the lower side ($-15H < z < -4H$) for all runs.
    Black contour shows the region where the accretion rate in the high-altitude region exceeds the averaged accretion rate in the disk region ($|z| \leq 4H$).
    The dotted line represents the gas motion that feels the half of the gravity force, $-GM/2r$, for the reference.
    Note that the cases without the Hall effect does not show any values in this plot.  
    \label{fig:haacc}
    }
\end{figure*}

To better quantify these variabilities, we present a space-time diagram of the net accretion rate in the lower disk surface ($-15H < z < -4H$), as shown in Figure \ref{fig:haacc}. We measure the accretion rate in broader regions to include the surface layers (Figure \ref{fig:H+_timevar}). In the \rFHp run, the accreting material first appears at outer radii ($\sim$ 6 au) and migrates inward at a high velocity that can be approximated by a particle in free fall under half the gravitational force. 
This process repeats in a quasi-periodic manner: each accretion episode lasts about 15 years, and the next one begins roughly 10 years after the previous episode ends.
As the clump reaches the inner region, the accretion rate becomes highly variable and may exceed the bulk accretion rate, suggesting that episodic surface accretion can significantly enhance variability in the stellar accretion rate.

We also find that when the Hall effect is not included, the time variability in the accretion rate of the disk surface is not observed. In addition, when the irradiation opacity is increased, the amplitude of the variabilities is damped.
Based on the phenomenology above, we next investigate the underlying physical mechanisms.

\subsubsection{Physical mechanisms}


We examine the physical mechanisms that govern the episodic accretion by focusing on clumps originating around $R \sim 5$ au in the \rFHp run (Figure \ref{fig:H+_timevar}).
The first row ($t = 306$ yr) illustrates a shadowed phase, where clumps in the inner region ($R \sim 2$ au) block direct optical and XUV irradiation from the central star.  
The optical shadow lowers the disk temperature around $R \sim 5$ au. 
The gas falls into the disk in the shadowed region, instead of launching outflows. Force examination indicates that
a steep pressure gradient between the irradiated and shadowed regions exerts a strong downward force.
Furthermore, in the XUV shadow, there is strong ambipolar diffusion around the current layer (in the shadow), which weakens the toroidal field, reducing magnetic pressure. These processes combined lead to the downward flow. The wind is still launched, but from much higher altitude under stellar illumination.


Meanwhile, in the inner irradiated regions,
the surface is locally heated up,
which assists the launching of thermal outflows ($t \sim 306$ yr).
The clump accretes inward by torque exerted by flipped magnetic fields, while gets reduced in size due to the outflow mass loss.
Once the clump reaches the inner buffer zone ($r < 0.8$ au), it dissipates numerically while in reality it is expected to quickly accrete to the central star.

The next phase is a wind-launching phase ($t \sim 313$ yr).
After the dissipation of the inner clump, the disk surface at $R \sim 4$--7 au becomes exposed to irradiation.
This leads to rapid expansion and reverts the inflow into an outflow, driven by a combination of thermal pressure and magnetic pressure gradinents. In the meantime, FUV also penetrates deeper, leading to better coupling between gas and magnetic fields ($\Am$ is brought above unity).

Subsequently, the uplifted gas accretes inward (an accreting phase; $t \sim 318$ yr).
We first note that the accretion flow is always present near the location where toroidal field flips in the disk surface, as generally resulting from the magnetic torque (proportional to $B_z \d B_\phi/\dz$).
As ambipolar diffusion at this location becomes ineffective, the poloidal magnetic fields get dragged by the accreting gas, stretched by Keplerian rotation, and drive further accretion. This process is exactly analogous to the development of the MRI channel modes, leading to efficient local angular momentum transport, causing fast accretion seen in Figure \ref{fig:haacc}. This locally rapid accretion further results in {\it local} mass accumulation, and hence the formation of an accretion clump. This clump is also
irradiated by XUV and optical radiation, thus maintaining the status of being warm and well coupled with the magnetic field. It also starts to shadow the regions further out.

Finally, as the accreting clump reaches the disk inner region,
further stretching of the poloidal fields leads to magnetic reconnection,
breaking the clump into multiple smaller clumps of magnetic islands (analogous to the break up of the MRI channel flows). The individual clumps in the accretion layer generate irradiated and shadowed regions, fragmenting the layer into clumpy structures, while being accreted toward the central star. After these clumps accrete and dissipate, the next winds emerge.

In the \rFHn run, similar accretion mechanisms operate.
The Hall effect damps horizontal magnetic fields in the disk, which also makes the current layer of the field flip at higher altitudes, although the effect is not significant.
Furthermore, surface heating by optical irradiation can lift the current layer, reducing its gas density.
This reduction triggers the rapid accretion by the stretched poloidal fields, once the ambipolar diffusion at the flip weakens, as in the \rFHp run. 
These processes repeat in cycles, resulting in episodic accretion.

On the other hand, in the runs without the Hall effect, time variations in the accretion rate at high altitudes are not observed (Figure \ref{fig:haacc}).
This is primarily because the strong current layer where the field flip is not elevated as in the cases with the Hall effect, and resides at deeper altitudes where ${\rm Am}$ is small (i.e., $\Am \lesssim 1$).


This explanation also aligns with the higher $\kapirr$ cases: episodic accretion still occurs, but the amplitude of variability is reduced (Figure \ref{fig:haacc}). 
The higher irradiation opacity leads to a narrower irradiated region because the irradiation
column is smaller at higher altitude.
Thus,
the column participating in episodic accretion is reduced accordingly.

In summary, episodic accretion in our simulation occurs when the Hall effect forces the current layer to high altitudes, enabling interaction with irradiation fronts. 
Irradiation may elevate the current layer, which triggers runaway accretion by radially stretched magnetic fields and form accretion clumps. 
Subsequently, the accretion clumps cast optical and XUV shadows in outer regions, creating a quiescent phase until their dispersal.

\section{Discussion}
\label{sec:discussion}

\subsection{Comparison with Previous Temperature Models of Irradiation Heating}
\label{ssec:comp-model}

\begin{figure}
    \centering
   \includegraphics[width=0.96\linewidth]{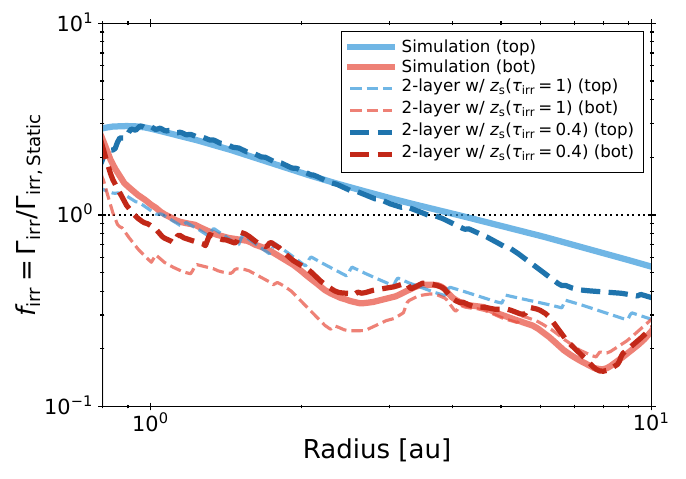}
    \caption{
    Same as Figure \ref{fig:firr}, but for \rFHp, with $f_{\rm irr}$ shown separately for the top and bottom surfaces. The two-layer model with a surface optical depth of $\tauirr$ = 0.4 is also plotted.
    \label{fig:firrH+}
    }
\end{figure}


We have compared the simulation temperature with the Static model (see Section~\ref{sssec:static-model}), where the density structure is unaffected by the disk wind.
We have shown that the winds push up the irradiation front, thereby enhancing irradiation heating.
One of the classical models for irradiation heating is the two-layer model by \citetalias{Chiang1997Spectral-Energy}. We describe our implementation of the two-layer model in Appendix~\ref{app:anal}, and confirm that under a hydrostatic density profile, the two-layer model reasonably reproduces the temperature based on \citetalias{Hubeny1990Vertical-struct} and a Monte Carlo simulation. In Figure \ref{fig:firr}, we further examine an modified two-layer model with the irradiation surface ($\tauirr = 1$) estimated from our simulations. We see that the heating rates in the updated two-layer model show a similar trend to those in the simulation, though they are still systematically smaller, especially in the \rFHp run. We attribute the deviation to the fact that irradiation heating is not localized at the irradiation surface ($\tauirr = 1$), but is distributed over a range of heights above the irradiation surface. In the \rFHp run, we find that the two-layer model with the $\tauirr \approx 0.4$ surface gives a temperature that matches well with the simulation in the inner region (Figure \ref{fig:firrH+}).
Therefore, we conclude that irradiation heating can be modeled using the two-layer model with the effective irradiation surface above the $\tauirr = 1$ surface.

We also compare our temperature structures with the recent radiative MHD simulation by \citet{Gressel2020Global-Hydromag}.
Their simulations exhibit higher disk temperatures than ours even for lower stellar luminosity and the same opacities. We notice that they adopted a higher initial disk temperature (with $H/R\sim0.05$), and the initial temperature is largely preserved during their simulations of $\sim100$ yrs (O. Gressel, private communication).
In our simulations, the midplane temperature achieves an equilibrium state within $\sim$30 yr, which is almost consistent with the estimation for the thermal relaxation timescale \citep[10--100 yr; e.g.,][]{Lyra2019The-Initial-Con}. 
The reason of the discrepancy is not entirely clear, but their work mainly focused on the thermodynamics in the wind region, which we treated very roughly.

\subsection{Snowline location}
\label{ssec:sl}


\begin{figure}
    \centering
   \includegraphics[width=0.95\linewidth]{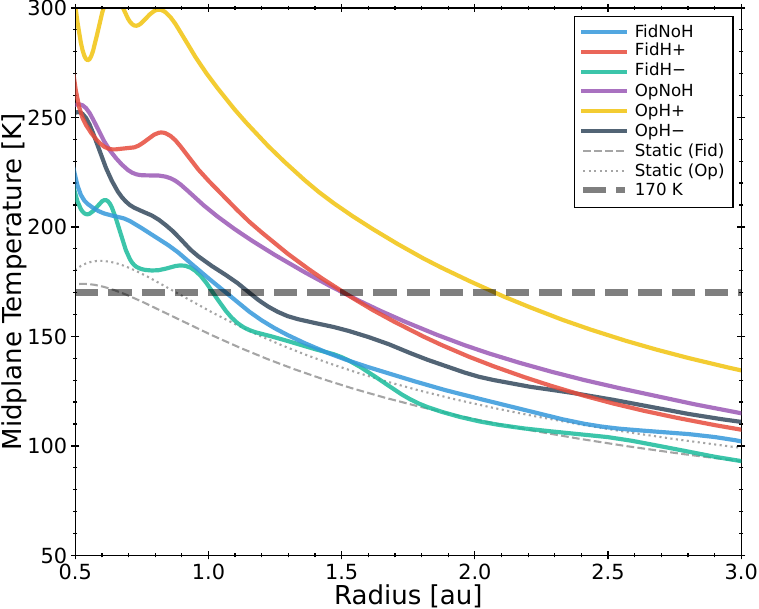}
    \caption{
    Midplane temperature profile around the snowline ($T=170$ K), with the time average over 200 yr.
    The colored solid lines represent the result of the simulation.
    The gray thin lines represent the temperature of the Static model for comparison.
    \label{fig:Tmid_sl}
    }
\end{figure}


Our simulations show that even with the same stellar luminosity ($L = 2.7 \Lsun$), gas surface density, and similar accretion rates (driven by the disk winds), the temperature profile in the bulk disk can vary by as large as a factor of two. This variation can strongly affect the location of the water snowline\footnote{We did not include physical processes around the snowline (e.g., dust transport and latent heat effects), which can affect the thermal structure and introduce additional dynamical complexity \citep{Wang2025Solving-for-the}.}. Figure \ref{fig:Tmid_sl} summarizes the midplane temperature around the snowline for all our runs and models.
In the Static model, the snowline location for $\kapirr = 1 $ \cmcmg is $\sim 0.6 $ au, and that for $\kapirr = 10 $ \cmcmg is 0.8 au. In our simulations, the snowline is located further outwards due to the enhanced irradiation heating, and can reach a maximum of 2.1 au in the \rOHp run. 

As a result of the higher disk temperature, the inward migration of the snowline (as stellar luminosity decreases over time) could be delayed in magnetized PPDs.
In \citet{Mori2021Evolution-of-th}, we found that the snowline arrives at 1 au at $t = 0.6$ Myr based on the results of local simulations.
Given our global simulation results, we find that the snowline arrival time at 1 au would be $\sim$ 2 Myr if we simply assume that the snowline location in the magnetized disks is two times farther from the center compared to the Static model.
The delayed arrival of the snowline at 1 au likely affects the water supply in the terrestrial planet forming region strongly \citep[e.g.,][]{Mori2021Evolution-of-th,Kondo2023The-Roles-of-Du}.



\subsection{Observational implications}
\label{ssec:comp-obs}

In this subsection, we discuss implications of our results on three aspects of disk observations.



%
%


First, our simulation results suggest many disk outflows are likely asymmetric in nature, thanks to the Hall effect. In our simulations, the density contrast on the two sides of the disk launched from the inner few au can be as large as $\sim100$, with local outflow rate differing by more than one order of magnitude (particularly the case with the aligned field polarity).  While constraining the wind kinematics observationally can be subject to major uncertainties, there has been increasing evidence over the recent years that reveals similar top-bottom asymmetry in some nearly edge-on disks \citep[e.g.,][]{Fernandez-Lopez2013Multiple-Monopo, Garufi2020ALMA-chemical-s, de-Valon2022Modeling-the-CO}. For instance, the ALMA $^{12}$CO observation of the HH 30 edge-on disk \citep{Louvet2018The-HH30-edge-o} only detected a one-sided outflow. JWST has greatly expanded the sample of disk winds seen in both continuum and gas line emissions in the near and mid-infrared. Many of the edge-on systems again exhibit top-bottom asymmetry 
\citep[e.g.,][]{Arulanantham2024JWST-MIRI-MRS-I,Delabrosse2024JWST-study-of-t,Villenave2024JWST-Imaging-of,Birney2024A-kinematical-s,Pascucci2025The-nested-morp}.
Interestingly, some of the asymmetric outflows have been suggested to originate from within a few au \citep[e.g.,][]{Louvet2018The-HH30-edge-o,de-Valon2020ALMA-reveals-a-,Booth2021Molecules-with-}, which might suggest that the Hall effect, which dominates in such inner regions, plays a key role in shaping the asymmetric outflows.
We note that from our findings, the side with weak and tenuous outflows corresponds to where the strong (and potentially episodic) surface accretion layer resides.
In addition, on the side with less density, the magnetic lever arm of the wind tends to be large, which is observed in HL Tau outflow \citep{Bacciotti2025ALMA-chemical-s}.
Therefore, asymmetric outflows can be considered as evidence for magnetically driven accretion in PPDs, and our finding further offers a guide to search for such fast surface accretion flows \cite[e.g.][]{Najita2021High-Resolution}.




Second, the dense side of the wind may effectively shield stellar UV radiation, allowing molecules, such as water vapor in the wind, to survive from photodissociation. 
This mechanism has been proposed for the Class 0/I disks \citep{Panoglou2012Molecule-surviv}, while our study shows that this also applies to Class II disks. 
Since the Spitzer time \citep{Carr2008Organic-Molecul}, water molecules have been routinely detected. 
Thanks to JWST, and with dedicated programs such as MINDS \citep{Henning2024MINDS:-The-JWST,Kamp2023The-chemical-in}, JDISC \citep{Pontoppidan2024High-contrast-J} and JOYS \citep{van-Gelder2024JOYS:-Mid-infra}, it offers a remarkable opportunity to characterize the volatile inventory in the warm inner disk surface layers \citep{Grant2023MINDS.-The-Dete,Grant2024MINDS:-A-multi-,Perotti2023Water-in-the-te,Temmink2024MINDS:-The-DR-T,Temmink2024aMINDS:-The-DR-T,Romero-Mirza2024Retrieval-of-Th,Romero-Mirza2024JWST-MIRI-Spect,Schwarz2025MINDS.-JWST-MIR}. While water molecules self-shield stellar UV radiation \citep{Bethell2009Formation-and-S,Bosman2022aWater-UV-shield}, UV shielding by dust, potentially enhanced in the dense side of the disk wind, likely also plays an important role in some sources \citep{Gasman2023MINDS.-Abundant}.

Finally, our simulations suggest the presence of supersonic and episodic radial gas flows on the disk surface is likely, which may have two observational consequences. First, the episodic surface accretion flows have peak accretion rates comparable to the mean accretion rate, leading to accretion variability. Nevertheless, since our simulations do not model the $\lesssim1$au region, where a transition to the MRI-active zone likely buffers the materials and modulates the actual accretion rate onto the protostar \citep{Iwasaki2024Dynamics-Near-t}. We thus anticipate that such episodic surface accretion likely corresponds to ``routine" variability over timescales of $\sim10$ years observed in protostars \citep{Venuti2015UV-variability-,Sergison2020Characterizing-,Fischer2023Accretion-Varia}. Second, we note that evidence of fast and variable surface accretion flows has been reported in several sources through infrared spectroscopy of molecular lines \citep{Boogert2002High-Resolution, Zhang2015Dimming-and-CO-,Najita2021High-Resolution}. The reported flow velocity is of the order of a few km/s or more with temperatures of 250-500 K, carrying an accretion rate of $10^{-8}-10^{-7}\Msun$ yr$^{-1}$, consistent with our results scaled to $\sim$au scale. We thus anticipate that such surface accretion layers can likely be detected in more sources, especially when the disk orientation is favorable.

\subsection{Caveats and future prospects}

Our simulations and results are subject to a number of limitations and caveats. In terms of the included disk physics, we have simplified the treatment of XUV heating and the associated photochemistry, which could affect the kinematics and observable signatures \citep{Wang2019Global-Simulati,Gressel2020Global-Hydromag}, though we anticipate our treatment to at least qualitatively capture the expected behavior. Our implementation of radiation transport for the disk thermal emission relies on the two-stream approximation, and can be less accurate when the irradiation surface is high above the disk, calling for improvements. Moreover, we have adopted a constant dust opacity, whereas in reality, the opacity should strongly depend on dust transport and growth \citep{Kondo2023The-Roles-of-Du}.

In addition, our simulations are conducted in 2D. The laminar nature of the bulk disk suggests that our results likely carry over to 3D, whereas further work is needed to clarify the situation. Moreover, we have not yet broadly explored the parameter space, especially since we have fixed the poloidal field strength with $\beta_0=10^4$. It is expected that varying this disk magnetization parameter could also strongly affect several aspects of the results, especially on wind kinematics and variability, as partially explored in \citet{Bai2017Global-Simulati}. These aspects are left for future work.

Despite the presence of wave-like activities in the disk associated with time variability (see Section \ref{ssec:hall-irr-acc}), 
our simulations do not clearly show the development of hydrodynamic turbulence \citep[e.g.,][]{Lyra2019The-Initial-Con}.
This may be partly due to numerical limitations, such as 
insufficient resolution ($\dr / H \sim $ 0.3) which suppresses some instabilities
\citep[e.g., Convective overstability;][]{Lyra2014Convective-Over,Klahr2023Thermal-instabi,Klahr2024Thermal-barocli}, and the axisymmetric domain which cannot capture nonaxisymmetric modes \citep[e.g., Zombie vortex instability;][]{Marcus2016Zombie-Vortex-I}.
In addition, the physical conditions for instability are not satisfied.
For example, vertical shear instability in the inner region is likely suppressed because of long thermal relaxation timescales $t_{\rm relax}$, 
where $t_{\rm relax}$ at the midplane in our simulation is $t_{\rm relax} \Omega \sim 1 (r/\au)^{-2} (kH/10)^2 $  \citep{Malygin2014Mean-gas-opacit} with the perturbation wavenumber $kH \sim 10$ \citep[e.g.,][]{Stoll2014Vertical-shear-}.
Furthermore, magnetic tension in the ionized disk surface and outer regions also inhibits the growth of such instabilities \citep[e.g.,][]{Cui2020Global-Simulati,Latter2022The-vertical-sh}.
Further high-resolution MHD simulations are required.

Furthermore, the strong current in the accretion layers may cause anomalous magnetic diffusivities. Strong electric fields at the layer can heat charged particles \citep{Inutsuka2005Self-sustained-}, increasing their collisions with dust and reducing the ionization fraction, which in turn raises the diffusivities \citep{Okuzumi2015The-Nonlinear-O,Mori2016Electron-Heatin,Mori2017Electron-Heatin,Okuzumi2019The-Generalized}. Plasma-scale instabilities induced by strong fields may also contribute to enhanced resistivity \citep{Hopkins2024Microphysical-R}. These effects could suppress the accretion, though a detailed analysis is left for future work.



\section{Summary}\label{sec:sum}


In this work, we have performed the global nonideal MHD simulations with radiation transport, focusing on the inner region ($R < 10$ au) of PPDs.
We have investigated the thermal structure of PPDs consistently with their dynamics.

Our main findings are as below:
\begin{itemize}
\item
We have developed the simplified radiation transport method for PPDs (Section~\ref{ssec:method-rt} and Appendix~\ref{app:Tre}), motivated by \citet{Xu2021Formation-and-e} and \citet{Gressel2020Global-Hydromag}, which enables us to calculate the temperature structure with low computational cost and reasonable accuracy (see Appendix~\ref{app:test} for comparison with Monte Carlo simulations)

\item
The simulations including the Hall effect with the positive field polarity produce the plane-asymmetric disk winds (Figures \ref{fig:Hp-map} and \ref{fig:low-dens-wind}).
This asymmetry is likely caused by the Hall effect (the HSI) which strongly amplifies the horizontal magnetic field, pushing a strong current layer (where the horizontal field flips with strong accretion flow) to higher altitudes.
The Lorentz force focus the gas into the current layer, pushing
the wind base to even higher altitudes with substantially reduced wind density compared to the other side.


\item
Stellar radiation impacts the dynamics of the disk surface: episodic accretion occurs when the Hall effect forces the current layer to high altitudes, enabling interaction with XUV irradiation fronts.
The disk surface, once newly exposed to XUV radiation, first pushes away a thermal outflow. Meanwhile, with higher ionization, MRI channel flow-like activities are triggered near the strong current layer, leading to the formation of accreting clumps that cast shadows.
In the shadowed region, the absence of XUV heating and ionization leads to strong ambipolar diffusion and weak thermal pressure, which together suppress the outflow.
The cycle repeats itself after the clumps get accreted onto the central star and the disk becomes irradiated again.
In contrast, in runs without the Hall effect, the current layer resides at deeper altitudes, which does not drive episodic accretion.

\item
Accretion energy is released primarily as Joule heating, mainly due to the vertical gradient of the toroidal magnetic field (contribution from radial gradient is largely negligible). 
However, as Joule heating primarily occurs at high altitudes (above $\sim 2 H$), it usually contributes negligibly to the disk temperature compared to irradiation heating (Figure \ref{fig:temp-profs}).
Most of the released accretion energy is lost to the disk wind.

\item
Irradiation heating can be enhanced by the wind, as it elevates the thermal irradiation front which intercepts more radiation. This effect depends on the opacity to the stellar optical radiation (Figure \ref{fig:all_temps}), and 
can be understood by a two-layer model.
In our \rFHp run, with a high-density wind, the total heating rate at $r = $ 1 au is 4 times higher than in the hydrostatic equilibrium disk model, and the location of the snowline can be pushed substantially outwards (from $\sim0.6$ to $\sim2$ au).
In this case, the effective irradiation front is higher than the $\tauirr = 1$ surface because of the broader irradiation absorption layer in the wind.
\end{itemize}


Our simulation results offer insights into several aspects of recent PPD observations, including the one-sided outflows observed in HH 30 \citep[e.g.,][]{Louvet2018The-HH30-edge-o,Lopez-Vazquez2024Multiple-Shells}, trans-sonic and episodic surface accretion flow \citep{Najita2021High-Resolution}, and the survival of water vapor in the PDS 70 system \citep{Perotti2023Water-in-the-te}. Future observations (JWST, ngVLA, etc.) are expected to place stronger constraints on the distribution of solids and major chemical species, together with more delicate kinematic information in protoplanetary disks, which are essential to test the generality of such phenomena.



Our simulations are still subject to major uncertainties, particularly associated with the treatment of dust opacities, and the treatment of XUV irradiation. In the future, it is highly desirable to fully capture the evaporative wind physics so as to study its interplay with magnetized disk winds and assess the robustness of our simulation results.

\begin{acknowledgments}
We thank Oliver Gressel for his detailed discussion on the thermal structure of the disk and for providing his data. 
We also thank Takahiro Ueda, Mario Flock, Gregory Herczeg, and Yuhiko Aoyama for fruitful discussions.
This work is supported by a joint Tsinghua-Tohoku collaboration initiative, by Japan Society for the Promotion of Science KAKENHI grant Nos. JP21H04487, JP21H04495, JP21J00086, 22KJ0155, JP22KK0043 and 22K14081, and by National Science Foundation of China under grant No. 12233004, 12325304, 12342501.

\end{acknowledgments}

\appendix 

\section{Derivation of the time-averaged accretion rate}
\label{app:acc-rate}

In this appendix, we derive the time-averaged accretion rate from the angular momentum equation in axisymmetric spherical coordinates, following \citet{Aoyama2023Three-dimension}. 
As they did not explicitly clarify the time averaging, we here carefully take a time average to reduce time-derivative terms. 
The specific angular momentum is decomposed into the mean motion and residuals: 
$j = j_0 + \delta j$, where $j_0$ represents the specific angular momentum averaged over the time and $\theta$-direction (Equation \ref{eq:j0}). 
The rotational velocity is also expressed as $v_\phi = v_0 + \delta v_\phi$, where $v_0 = j_0 / R$.
The angular momentum equation is,
\begin{eqnarray}\label{eq:app-accrate-1}
	\frac{\partial (\rho j)}{\partial t} 
	+ \nablap \cdot \pr{ \rho \bm{v} j  - R \frac{\bm{B} B_{\phi}}{4\pi}  } = 0, 
\end{eqnarray}
where the subscript ``p'' denotes the poloidal component of the vectors, and 
we define the divergence operator (for a vector $\bm{A}$) on the poloidal plane as,
\begin{equation}
	\nablap \cdot \bm{A} = \frac{1}{r^2} \pd{(r^2 A_r )}{r}  + \frac{1}{R}\frac{\partial ( \sint  A_\theta) }{\partial \theta}
\end{equation}
The term on $j_0$ can be transformed as: 
\begin{equation}\label{eq:app-accrate-2}
\nablap \cdot \pr{ \rho \bm{v} j_0} 
=  \rho v_r \djdr -  j_0 \pd{\rho}{t} ,
%
\end{equation}
where 
$j_0' = \d j_0/\dr $, and we use the continuity equation. 
Note that $j_0$ depends only on $r$.
Using Equation \ref{eq:app-accrate-2}, 
Equation (\ref{eq:app-accrate-1}) becomes:
\begin{equation}\label{eq:app-accrate-3}
\frac{\partial (\rho \delta j )}{\partial t}  + \rho v_r \djdr 
+ \nablap \cdot \pr{ R  \bm{T}_{\phi} } = 0 ,
\end{equation}
where $ \bm{T}_{\phi}         = \rho \bm{v}_{\rm p} \delta v_\phi- \bm{B}_{\rm p} B_{\phi} /4 \pi$ .
%
We integrate Equation (\ref{eq:app-accrate-3}) over the disk volume as in \citet{Aoyama2023Three-dimension}:
\begin{equation}\label{eq:app-accrate-4}
\begin{split}
&\frac{\partial }{\partial t} \pr{  \int_{\rm disk} 2\pi R \rho \delta j r\d\theta }
+ \djdr \int_{\rm disk}  2\pi R \rho v_r  r\d\theta \\
&+ \frac{\partial}{\partial r } \pr{\int_{\rm disk}  2\pi R^2  T_{r\phi} r \d\theta }
+  \subtr{ 2 \pi   R^2 T_{\theta\phi}  } = 0 .
\end{split}
\end{equation}
%
Then, we take a time average to eliminate the time-derivative term.
We obtain the time-averaged accretion rate:
\begin{align}
\label{eq:app-accrate-5}
\avt{\dot{M}_{\rm acc}} &\equiv  \avt{ \int_{\rm disk}  2\pi R \rho v_r  r\d\theta}  \\
&\approx -  \frac{1}{\djdr } 
\left\{
 \avt{ \frac{\partial}{\partial r } \int_{\rm disk}  2\pi R^2 T_{r\phi} r \d\theta }
+ \avt{ \subtr{ 2 \pi    R^2 T_{\theta\phi}   } }
\right\} . \nonumber
\end{align}
This formula is applicable even if the integral range depends on time.

\section{Derivation of approximated energy balances}\label{app:ene-bug}

In this appendix, we derive the approximate balance of total energy change rates in spherical coordinates, as in \citet{Suzuki2016Evolution-of-pr}. 
We begin with the total energy equation (Equation (\ref{eq:ene-eq})) decomposing the rotational velocity into the mean motion ($v_0 = j_0 / R$) and residuals ($\delta v_\phi$), as in Appendix~\ref{app:acc-rate}.
The total energy equation is then described as,
\begin{align}
&\pd{E^{(0\text{--}2)} }{t} 
	+ \nablap \cdot \bm{F}^{(0\text{--}2)} 
	= q_{\rm rad},\label{eq:app-deren-1-a} \\
	 &~~E^{(0)} =  \frac{\rho v_0 ^2}{2} + \rho \Phi \equiv \rho \Phi_{\rm eff}  \, , ~~
	 E^{(1)} = \rho v_0 \delta v_\phi \, , ~~ \nonumber \\
	 &~~E^{(2)} = \frac{\rho \delta v_\phi ^2}{2} + \rho e + \frac{B^2}{8\pi}, \nonumber  \\
	&~~\bm{F}^{(0)} = \rho \bm{v}_{\rm p} \Phi_{\rm eff} \,,
	~~\bm{F}^{(1)} = \rho v_0 \delta v_\phi \bm{v}_{\rm p} - \frac{v_0 B_{\phi}}{4 \pi} \bm{B}_{\rm p} \equiv v_0 \bm{T}_\phi \, ,  \nonumber  \\
	&~~\bm{F}^{(2)} = \bm{v}_{\rm p} \pr{ E^{(2)} + P^*} 
				- \frac{\delta \bm{v}\cdot \bm{B}}{4 \pi} \bm{B}_{\rm p} + \bm{S}, \nonumber 
\end{align}
where the superscript in $E^{(*)}$ and $\bm{F}^{(*)}$ indicates the order of magnitude,
and the superscript $(0\text{--}2)$ denotes the sum of terms from the zeroth to second order.
To simplify the energy balance, we retain the terms that are similar to or larger than $O(\rho c_s^2 \Omega)$. 
For instance, the radial derivative on $F_{r}^{(2)}$ would be of the order of $O(\rho c_s^2 h \Omega)$, where the disk aspect ratio $h \ll1$, and thus be negligible.



Using Equation (\ref{eq:app-accrate-3}),
the radial derivative in $\nablap \cdot \bm{F}^{(0)} $ can be transformed into:
\begin{equation}\label{eq:app-deren-2}
\begin{split}
	&\frac{1}{r^2} \pd{}{r} \pr{ r^2 F_{r}^{(0)} } 
		=  \frac{\Phi_{\rm eff}}{r^2} \pd{(r^2 \rho v_r)}{r}   +  \rho v_r  \pd{ \Phi_{\rm eff}}{r} \\
	&	= - \frac{\Phi_{\rm eff} }{R} \pd{(\sint \rho v_\theta)}{\theta} 
		- \frac{1}{\djdr}  \pd{\Phi_{\rm eff}}{r} \nablap \cdot \pr{ R \bm{T}_\phi } \\
			&~~~~	- \pd{}{t} \pr{ \rho  \Phi_{\rm eff}  + \frac{\rho \delta j }{\djdr} \pd{ \Phi_{\rm eff}}{r} } ,
\end{split}
\end{equation}
where we also use the continuity equation. 
The first term in the RHS of Equation (\ref{eq:app-deren-2}) can be transformed as:
\begin{equation}\label{eq:app-deren-3}
\begin{split}
    &\frac{\Phi_{\rm eff} }{R} \pd{(\sint \rho v_\theta)}{\theta}\\
    &= \frac{1}{R}\pd{ \pr{ \sint  \rho v_\theta \Phi_{\rm eff} }}{\theta}  
    - \frac{\sint \rho v_\theta }{R} \pd{\Phi_{\rm eff}}{\theta}\\
    &= \frac{1}{R}\pd{ \pr{ \sint  \rho v_\theta \Phi_{\rm eff} }}{\theta}  
    + \frac{\rho v_\theta v_0^2 \cos\theta}{R} ,
  \end{split}
\end{equation}
where the divergence of $\bm{F}^{(1)}$ is expressed as,
\begin{equation}\label{eq:app-deren-4}
	\nablap \cdot \bm{F}^{(1)} 
	= \frac{v_0}{R} \, \nablap \cdot \pr{  R \bm{T}_\phi }   + R \bm{T}_\phi \cdot \nabla \pr{ \frac{v_0}{R}  } .
\end{equation}
By combining Equations (\ref{eq:app-deren-2}) and (\ref{eq:app-deren-3}), 
we have the term 
\begin{equation}
	\nablap \cdot \pr{  R \bm{T}_\phi } \cdot  \pr{ \frac{v_0}{R}  - \frac{1}{\djdr}  \pd{\Phi_{\rm eff}}{r} } 
	\approx O(\rho c_s^2 h \Omega),
\end{equation}
which is negligible because 
$ v_0/R  -  (\djdr)^{-1} \partial \Phi_{\rm eff}/\partial r $ is of the less order of $O(\Omega h^2)$.
Furthermore, in Equation (\ref{eq:app-deren-4}), the $\theta$ derivative of $v_0/R$ is of the order of $O(\Omega h /R)$, and thus the term is negligible.
Then, by adding $ \partial F_{\theta}^{(1)} / (R \partial \theta) $ to both sides of Equation (\ref{eq:app-deren-1-a}) and using Equations (\ref{eq:app-deren-2})--(\ref{eq:app-deren-4}), we obtain the approximate energy  balance equation:
\begin{equation}
\begin{split}\label{eq:app-deren-5}
&\pd{E^{(2)}}{t}
+ \frac{1}{R} \pd{}{\theta} \left[ \sint F_{\theta}^{(1\text{--}2)}\right] 
- \frac{\rho v_\theta v_0^2 \cos\theta}{R} \\
&\approx 
- R T_{r\phi} \pd{}{r} \pr{ \frac{v_0}{R}  }
+ \frac{\partial( \sint v_0 T_{\theta \phi})}{R\partial  \theta} 
+   q_{\rm rad} ,
\end{split}
\end{equation}
where we add $ \partial F_{\theta}^{(1)} /R \partial \theta  $ to both hand sides to obtain a similar formula as in \citet{Suzuki2016Evolution-of-pr},
and we note that $(\rho \delta j /\djdr) (\partial\Phi_{\rm eff}/\partial r ) = \rho \delta j ( v_0/ R + O(\Omega h^2) ) \approx  \rho \delta v_0 v_\phi $.

We integrate Equation (\ref{eq:app-deren-5}) with $\int_{\rm disk} \sint r \d \theta$, and then take the time average. Eventually, we obtain the approximate balance of the total energy change rate: 
\begin{equation}
\begin{split}
\label{eq:app-deren-7}
 & \avt{ \subtr{ \sint F_{\theta}^{(1\text{--}2)} } }
 - \avt{  \int_{\rm disk}  \frac{\rho v_\theta v_0^2 \cos\theta}{R}  \cdot  \sint r \d \theta }\\
  &\approx 
  - \avt{  \int_{\rm disk}  R T_{r \phi} \pd{}{r} \pr{ \frac{v_0}{R}  } \cdot  \sint r \d \theta  } \\
 & ~~~~ +  \avt{ \subtr{  \sint v_0 T_{\theta \phi}  } } 
 + \avt{  \int_{\rm disk}   q_{\rm rad} \sint r \d \theta } ,
\end{split}
\end{equation}
where 
the first and second terms in LHS denote the energy loss by the disk wind,
the first term in RHS denotes energy production by shearing motion,
the second term denotes the wind torque exerting the disk,
and the third term denotes net radiative heating.

Moreover, we then obtain the energy balance of the mechanical energy change rate.
The mechanical energy equation is obtained by the subtraction of the internal energy equation from the total energy equation.
The internal energy equation in axisymmetric disks is 
\begin{equation}
	\pd{(\rho e)}{t} + \nablap \cdot \pr{ \bm{v}_{\rm p} \rho e} =  - P \nablap \cdot \bm{v}_{\rm p} + q_{\rm rad} + q_{\rm Joule} .
\end{equation}
Integrating this equation along the $\theta$ direction and taking the time average, we obtain
\begin{equation}
\label{eq:app-deren-8}
\begin{split}
	 &\avt{ \subtr{  \sint    v_\theta \rho e  } }  
	\approx  
	- \avt{ \int_{\rm disk} P \pd{(\sint v_\theta)}{\theta} \d \theta} \\
	& + \avt{ \int_{\rm disk}   q_{\rm rad} \sint r \d \theta } 
	 + \avt{ \int_{\rm disk}   q_{\rm Joule} \sint r \d \theta } .
\end{split}
\end{equation}
Combining Equations (\ref{eq:app-deren-7}) and (\ref{eq:app-deren-8}), we obtain the energy balance of the mechanical energy change rate:
\begin{align}
\label{eq:app-deren-9}
    & \avt{ \subtr{ \sint \pr{ F_{\theta}^{(0\text{--}2)} \!\!\!- v_\theta \rho e }\! } }
        - \avt{ \int_{\rm disk}  \mspace{-10mu} \rho v_\theta v_0^2 \cos\theta  \d \theta } \nonumber \\    
    & +\avt{  \int_{\rm disk}    \mspace{-10mu} q_{\rm Joule} \sint r \d \theta } \nonumber  \\
    & \approx 
    - \avt{  \int_{\rm disk}  R T_{r\phi} \pd{}{r} \pr{ \frac{v_0}{R}  } \cdot  \sint r \d \theta  }
     + \avt{ \subtr{  \sint v_0 T_{\theta \phi}  } }  \nonumber  \\
     &~~~~ + \avt{  \int_{\rm disk} P \pd{(\sint v_\theta) }{\theta} \d \theta} .
\end{align}


\section{Derivation of the equation for the re-emission temperature}\label{app:Tre}

We describe here how to calculate the re-emitted radiation temperature, $\Tre$.
The radiation temperature is related to the mean radiation intensity $J$ by $ \sigma \Tre^4 =  \pi J $.
To obtain $J$, we solve the radiative transfer equation under the plane-parallel approximation. 
We solve $J$ along the $\theta$ direction due to the coordinate limitation, and do it for each radius.
Also, we calculate the emission and absorption of radiation from dust, based on the dust temperature.
This treatment of the three temperatures (gas, radiation, and dust) is necessary when XUV heating is included.
The XUV heating primarily increases the gas temperature, but not the dust temperature.
If the temperature of the XUV-heated gas is used in the calculation, the re-emitted radiation would be significantly higher than the dust radiation, leading to incorrect temperature structures.

We start with a radiative transfer equation for rays with a wave frequency $\nu$, at a colatitude $\theta$ on a radius $r$, 
\begin{eqnarray}\label{eq:rt-1}
   \frac{\mu}{ \rho \kappa_\nu r } \pd{I_\nu (\theta, \mu) }{ \theta } = - I_\nu(\theta, \mu) + S_\nu(\theta),
\end{eqnarray}
where $S_\nu$ is the source function, and 
 $\mu$ is the cosine of the angle between the ray and the normal direction of the plane.
Firstly, we assume a two-stream approximation where the rays have characteristic angles upward and downward, 
\begin{eqnarray}
	I_\nu (\theta, \mu) = \left\{ 
	\begin{array}{ll}
		I_\nu^+(\theta) \,\delta(\mu - \mu_0), & ~~( \mu > 0), \\
		I_\nu^-(\theta) \,\delta(\mu + \mu_0), & ~~( \mu < 0),
	\end{array} \right .
\end{eqnarray}
where $I_\nu^+$ and $I_\nu^-$ are the radiation intensity for rays propagating upward and downward, respectively,
and $\delta$ is the delta function.
The mean radiation intensity is described as: 
\begin{eqnarray}
 	J_\nu (\theta) = \frac{1}{2} \int_{-1}^1 I_\nu (\theta, \mu) \d \mu 
		       = \frac{ I_\nu^+ + I_\nu^- }{2} .
\end{eqnarray}
We take the cosine of the characteristic angle, $\mu_0$, to match with the Eddington approximation: $\mu_0 = 1 / \sqrt{ 3 }$.
Integrating \eqref{eq:rt-1} from $\mu = 0$ to $1$, we have the radiative transfer equation on $I_\nu^+$ 
\begin{eqnarray}\label{eq:rt-2}
   \frac{\mu_0}{ \rho \kappa_\nu r} \pd{I^+_\nu (\theta)}{ \theta } = - I^+_\nu(\theta) + S_\nu(\theta).
\end{eqnarray}

We then integrate Equation (\ref{eq:rt-2}) over the frequency, 
assuming local thermal equilibrium and neglecting scattering for simplicity.
The source function $S_\nu$ is given by the Planck function of the dust temperature because the radiation is predominantly emitted/absorbed by dust,  
\begin{eqnarray}\label{eq:source-func}
   \int  S_\nu \d\nu = \int B_\nu(T_{\rm dust}) \d \nu  = \sigma T_{\rm dust}^4/\pi .
\end{eqnarray}
Using Equation (\ref{eq:source-func}), we obtain the radiative transfer equation for $I^+$,
\begin{eqnarray}\label{eq:rt-4}
   \frac{\mu_0}{r}  \pd{I^+ (\theta) }{ \theta } = - \rho \kappa_{\rm disk} \pr{ I^+ - \frac{\sigma T_{\rm dust}^4}{\pi} } ,
\end{eqnarray}
where we assume that both the Planck-mean opacity and the intensity-weighted mean opacity are equal to $\kappa_{\rm disk}$.

To obtain the dust temperature $T_{\rm dust}$, we calculate the equilibrium of dust heating and cooling.
The dust is heated by absorbing stellar irradiation and re-emitted radiation, as well as
the collisions between dust and gas particles \citep{Yorke1996Photoevaporatio}, and is cooled by its own thermal emission,
\begin{equation}\label{eq:Tdust_1}	 
\begin{split}
	 &~4 \rho \kapirr \sigma T_{\rm irr,eq}^4 \exp\pr{-\tauirr}
	+ 4 \rho \kappa_{\rm disk} \sigma T_{\rm re}^4  \\
	&+ 4 \pi a^2 c_{\rm s}(T_{\rm gas}) n_{\rm g} n_{\rm d} k \pr{T_{\rm gas} - T_{\rm dust}} \\
	&~= 4 \rho \kappa_{\rm disk} \sigma T_{\rm dust}^4,
\end{split}
\end{equation}
where $n_{\rm g}$ and $n_{\rm d}$ are, respectively, the number density of the gas and dust particles, 
and $a$ is the radius of dust grains.
Dividing \eqref{eq:Tdust_1} by $4 \rho \kappa_{\rm disk} \sigma $, we obtain a quartic equation on $T_{\rm dust}$, 
\begin{equation}\label{eq:Tdust_2}
\begin{split}
	 &T_{\rm dust}^4 
	 - \frac{ \kappa_{\rm geom} }{ \kappa_{\rm disk} } \frac{c_{\rm s}^3(T_{\rm gas})  \rho}{\sigma T_{\rm gas} }  \pr{T_{\rm gas} - T_{\rm dust}} \\
	 &~~~ - \left[ \frac{\kapirr}{\kappa_{\rm disk}} T_{\rm irr, eq}^4 \exp\pr{- \tauirr} 
	+ T_{\rm re}^4 \right] = 0,
\end{split}
\end{equation}
where $ \kappa_{\rm geom} = \pi a^2 n_{\rm d} / \rho $ represents the opacity due to the geometrical cross-section of dust grains. 
For simplicity, we assume $\kappa_{\rm geom}/\kappa_{\rm disk}  = 1$, though this ratio should be consistent with dust properties and abundance as well as the opacities.
We update $T_{\rm dust}$ by solving Equation (\ref{eq:Tdust_2}) using the Newton-Raphson method every timestep before the calculation of $\Tre $.
Here, we use $T_{\rm gas}$ and $\Tre$ at the previous timestep before updating $\Tre$.  

Integrating Equation (\ref{eq:rt-4}) along both upward and downward $\theta$ directions for all radii, we obtain the spatial distribution of $I^+$ and $I^-$.
The integration domain is for $\theta = [\pi/6, 5\pi/6]$, with the initial value of the integration taken to be zero.
From the $I^\pm$ profiles, we calculate the radiation temperature as $\Tre^4 =  \pi (I^+ + I^-)/(2\sigma)$.

\begin{figure}
    \centering
   \includegraphics[width=0.86\linewidth]{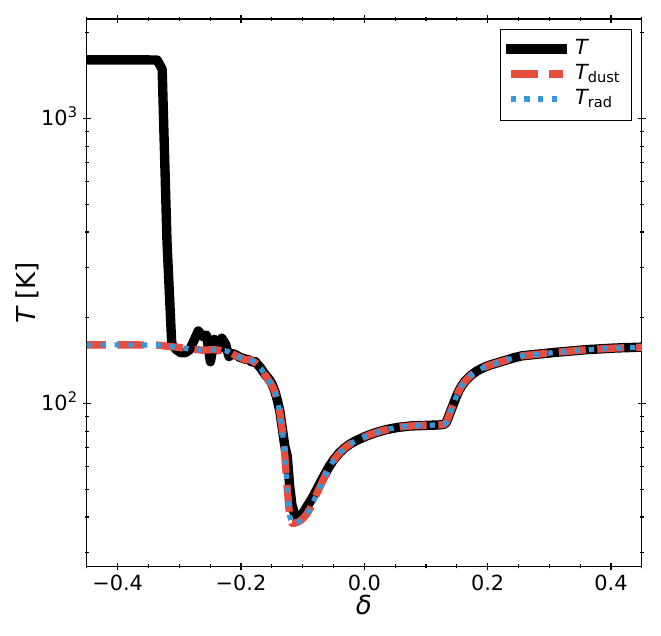}
    \caption{
    Vertical temperature profile at $r=5$ au and $t = 365$ in the \rFHp run.
    The solid line indicates the gas temperature, 
    dashed line the dust temperature used in the source term,
    and dotted line the radiation temperature including the stellar optical irradiation and infrared re-emitted radiation.  
    \label{fig:three-temp}
    }
\end{figure}

We show the vertical profile of the three temperatures, i.e., gas ($T$), radiation (infrared re-radiation; $T_{\rm re}$), and dust temperature ($T_{\rm dust}$) in the \rFHp run ($t = 365$ yr; $r = 5 $ au) in Figure \ref{fig:three-temp}. 
%
%
%
%
The three temperatures are coincident below the XUV-heated region.
Without XUV heating, the dust and gas temperature converge into the radiation temperature. 
In the XUV-heated region, the gas temperature deviates from the dust temperature. 
At a high altitude, collisional energy transfer between the gas and dust is not effective, and thus the dust temperature is still in equilibrium with the radiation temperature.

\section{Tests for the radiative transfer}
\label{app:test}

\begin{figure*}
    \centering
   \includegraphics[width=0.92\linewidth]{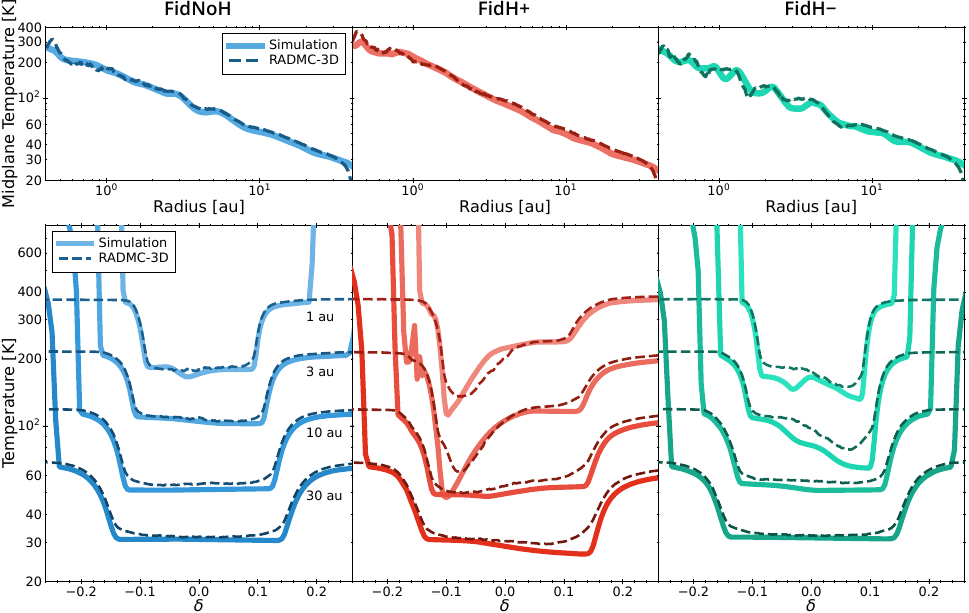}
    \caption{
    Comparison of the simulation temperatures ($t = 365$ yr) with those calculated from RADMC-3D.
    Each column represents the simulation runs: \rFNH (left), \rFHp (middle), and \rFHn (right).   
    The upper panels display the midplane temperature profiles, while the lower panels show the vertical profile around the midplane at $r = 1, 3, 10, 30$ au.
    \label{fig:radmc-test}
    }
\end{figure*}

In our simulations, re-emitted radiation propagates along the $\theta$ direction in optically thin regions, and we adopt the Eddington approximation, which is valid for isotropic radiation fields. It is thus necessary to
validate our approximate radiative transfer method by comparing the temperature profiles obtained from more realistic calculations.
%
%

In doing so, we carry out Monte-Carlo radiative transfer calculations using the RADMC-3D code \citep{Dullemond2012RADMC-3D:-A-mul}.
We use the density structure from our radiative MHD simulation at $t=365$ yr. 
We place a point light source with the same luminosity ($L = 2.7 L_\odot$) as that used in the radiative MHD simulation at the center.
The test operates under the assumption of local thermodynamic equilibrium, while disregarding scattering effects.
To make fair comparison, we adopt a wavelength-independent opacity,
taken to be constant of 1 \cmcmg to RADMC-3D, and compare the results with the simulation using $\kapirr = 1$ \cmcmg (\rFNH, \rFHp, and \rFHn), where the opacities for visible and infrared light are the same. 
We disregard opacity differences in FUV, as they have no impact on the temperature distribution near the disk.
Figure \ref{fig:radmc-test} presents a comparison of temperature distributions between our simple radiation transfer method and RADMC-3D.
The figure illustrates both the radial temperature distribution along the midplane and the vertical temperature distribution at specific radii ($r = 1, 3, 10, 30$ au).
Overall, the temperature distributions in our simulation closely match those from RADMC-3D.
The significant temperature jump observed at the inner boundary in RADMC-3D is due to the model assumption, specifically, its neglect of the optical depth inside the inner radial boundary.

One may notice the discrepancies seen in the vertical temperature distribution, which is just below the irradiation front.
This would be due to the Eddington approximation in the low-density regions.
Note that discrepancies near the midplane, particularly in the \rFHn run, are likely due to temporal variations, which cannot be captured by RADMC-3D.
Nevertheless, these discrepancies are within $\sim$ 10 \%,
suggesting that our radiative transfer method calculates the disk's temperature structure with sufficient accuracy.

\section{Analytical Profiles of Irradiation-heated Temperature Model }
\label{app:anal}

\begin{figure*}
    \centering
   \includegraphics[width=1\linewidth]{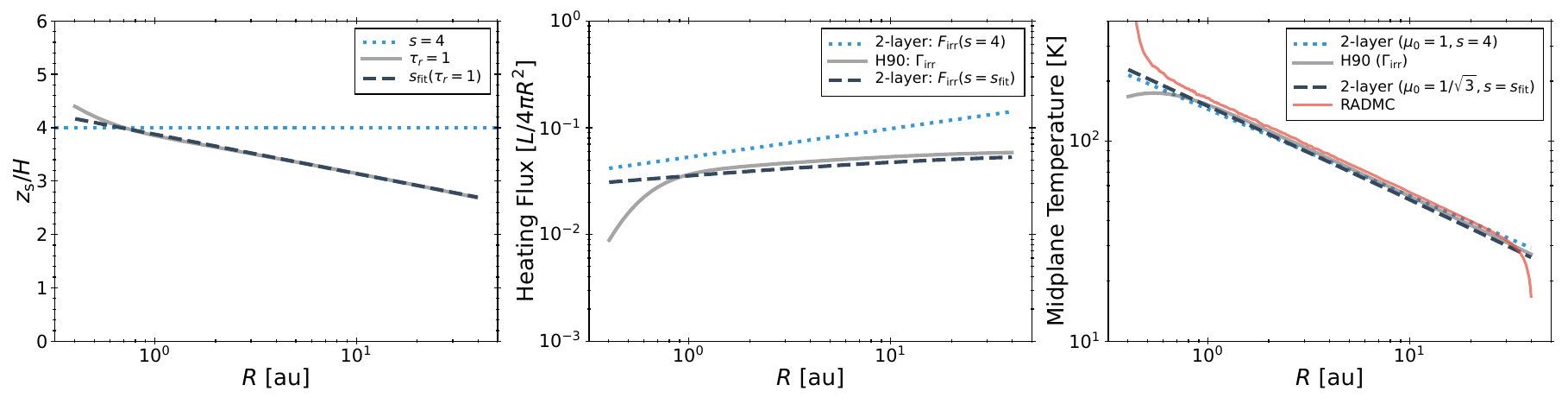}
    \caption{
    	Comparison of the temperature models of the passively heated disks where 
	the density structure is in vertically-isothermal hydrostatic equilibrium.
	The models are 
		the \citetalias{Chiang1997Spectral-Energy} two-layer model (dotted), 
		\citetalias{Hubeny1990Vertical-struct} model (solid),
		and modified two-layer model (dashed).
	Left: altitude of the irradiation front, normalized by the gas scale height.
    	Middle: irradiation heating rate per unit area, 
		normalized by the stellar radiation flux in the optically thin limit.
    	Right: midplane temperature profiles, with the red line indicating RADMC-3D calculation.
    \label{fig:app-tpanels}
    }
\end{figure*}

Here, we examine the temperature models of passively heated disks.
One of the classical models is the two-layer disk model \citepalias{Chiang1997Spectral-Energy},
where the irradiation energy is assumed to be absorbed within a narrow layer on the disk surface.
In addition, the temperature formula by \citetalias{Hubeny1990Vertical-struct} has also been used,
which provides the vertical temperature profile based on given density and heating rate profiles.
We aim to clarify the relationship between these models and compare them with a Monte Carlo simulation.

In the two-layer model, the midplane temperature in the optically thick region is described as,
\begin{equation}\label{eq:app-Tirr-CG}
	T_{\rm irr} = \pr{ \frac{F_{\rm irr} }{8 \sigma}  }^{1/4} ,  ~~ 
	F_{\rm irr} = \frac{L \alpha_{\rm graz} }{ 2 \pi R^2 } ,  ~~
	\alpha_{\rm graz} = R \frac{\d}{\d R} \pr{ s \frac{H}{R} } ,
\end{equation}
where $s \equiv z_{\rm s}/H$ is the irradiated altitude $z_{\rm s}$ normalized by the scale height $H$ and $s$ is often assumed to be 4.
The energy flux $F_{\rm irr}$ absorbed at the irradiated surfaces (top and bottom) is determined by the grazing angle $\alpha_{\rm graz}$ to the surface.
Then, half of $F_{\rm irr}$ propagates to the midplane as the re-emitted radiation.
The re-emitted radiation is modeled using the linear two-stream approximation (i.e., the cosine of the characteristic angle $\mu_0 = \pm 1$).
Indeed, we see that half of the total flux ($I^+ + I^-$) travels toward the midplane in Figure \ref{fig:Hp-J}.

In contrast, the \citetalias{Hubeny1990Vertical-struct} formula (Equation (\ref{eq:texp}) with symmetric structures) is based on the spatial distribution of a heating rate. 
Under the irradiation heating rate (Equation (\ref{eq:qirr})), the midplane temperature is described by 
\begin{equation}\label{eq:app-Tirr-Hubeny}
	T_{\rm irr} = \pr{ \frac{\sqrt{3} \Gamma_{\rm irr}}{8\sigma}}^{1/4}, ~~\Gamma_{\rm irr} = \int_{\rm disk} q_{\rm irr} r\dth
\end{equation}
where $\Gamma_{\rm irr}$ is the vertical integral of the heating rate.
Note that any vertical distribution of the heating rate is applicable. 
Here we calculate the irradiation heating rate from the radial column density based on the global density structure.

One of the differences between the models is the irradiation altitude, as shown in the left panel of Figure \ref{fig:app-tpanels}. 
We assume the same density structure as the Static model (see Section~\ref{sssec:static-model}). In this case, while \citetalias{Chiang1997Spectral-Energy} assumes $s = 4$, $s$ at the altitude where $\tau_r = 1$ varies from 4 to 3 from 1--10 au.
In addition, the heating energy rate per unit area in \citetalias{Chiang1997Spectral-Energy} is also higher than \citetalias{Hubeny1990Vertical-struct} (see the middle panel).

Is this due to differences in how to calculate the heating flux?
First, to see the influence of the $s$ distribution, we plot $F_{\rm irr}$ with the two-layer model that uses $s = s_{\rm fit}$, where $s_{\rm fit}$ is fitted to be the altitude of $\tau_r = 1$ (see dashed lines in the left and middle panels).
We obtain the fitting formula $ s_{\rm fit} = a_{\rm fit} \log_{10}(R/{\rm au}) + b_{\rm fit} $ with $a_{\rm ft} = -0.740$ and $b_{\rm fit} = 3.881$. 
Clearly, the radial slope of $s$ has significant impacts on the grazing angle. 
Decomposing $\alpha_{\rm graz}$ into terms that depends on the derivative $s$ and does not, 
\begin{equation}
	\alpha_{\rm graz} = s \frac{H}{R} \left[ \frac{\d \ln (H/R)}{ \d \ln R}  + \frac{H}{s} \frac{\d s}{\d \ln R}  \right] ,
\end{equation}
we find that the second term is $\sim -0.3$ to $-0.4$ times the first term.
Therefore, $\alpha_{\rm graz}$ is overestimated when assuming constant $s$, which then leads to the higher heating flux in the two-layer model with constant $s$.

Although the heating fluxes differ, the temperatures in the two models are well consistent in this disk structure (see the right panel in Figure \ref{fig:app-tpanels}). This is because the cooling rate adopted in \citetalias{Chiang1997Spectral-Energy} is effectively $2 \pi B_{\rm bb}$ (where $B_{\rm bb} = \sigma T_{\rm }^4 / \pi$ is the black-body radiation intensity), which is larger than the more accurate values of $2 \pi B_{\rm bb} /\sqrt{3}$ used in \citetalias{Hubeny1990Vertical-struct}.
Therefore, coincidentally, the midplane temperature becomes comparable between the two models.
Adopting the two-layer model with $s_{\rm fit}$, we further see that it well reproduces midplane temperature and its radial temperature slope obtained from the more realistic RADMC-3D calculations.

\bibliography{mybib_with_doi}
\bibliographystyle{aasjournalv7}

\end{document}